\documentclass[aps,prc,twocolumn,superscriptaddress,showpacs,floatfix,nofootinbib]{revtex4}

\usepackage{amsmath,graphicx}
\usepackage[colorlinks=true,linktocpage=true,linkcolor=blue,citecolor=blue]{hyperref}
\usepackage[usenames,dvipsnames]{color}
\usepackage{float}

\begin{document}

\title{Relativistic third-order viscous corrections to the entropy four-current from kinetic theory}

\author{Chandrodoy Chattopadhyay}
\affiliation{Department of Nuclear and Atomic Physics, Tata Institute of Fundamental Research, Homi Bhabha Road, Mumbai 400005, India}
\author{Amaresh Jaiswal}
\affiliation{GSI, Helmholtzzentrum f\"ur Schwerionenforschung, Planckstrasse 1, D-64291 Darmstadt, Germany}
\author{Subrata Pal}
\affiliation{Department of Nuclear and Atomic Physics, Tata Institute of Fundamental Research, Homi Bhabha Road, Mumbai 400005, India}
\author{Radoslaw Ryblewski}
\affiliation {The H. Niewodnicza\'nski Institute of Nuclear Physics, Polish Academy of Sciences, PL-31342 Krak\'ow, Poland}

\date{\today}

\begin{abstract}

By employing a Chapman-Enskog like iterative solution of the 
Boltzmann equation in relaxation-time approximation, we derive a new 
expression for the entropy four-current up to third order in 
gradient expansion. We show that unlike second-order and third-order 
entropy four-current obtained using Grad's method, there is a 
non-vanishing entropy flux in the present third-order expression. We 
further quantify the effect of the higher-order entropy density in 
the case of boost-invariant one-dimensional longitudinal expansion 
of a system. We demonstrate that the results obtained using 
third-order evolution equation for shear stress tensor, derived by 
employing the method of Chapman-Enskog expansion, show better 
agreement with the exact solution of the Boltzmann equation as well 
as with the parton cascade BAMPS, as compared to those obtained 
using the third-order equations from the method of Grad's 14-moment 
approximation.

\end{abstract}



\maketitle

\section{Introduction}

Study of the space-time evolution and non-equilibrium properties of 
hot and dense matter produced in high-energy heavy-ion collisions, 
within the framework of relativistic viscous hydrodynamics, has 
gained widespread interest; see Ref. \cite{Heinz:2013th} for a 
recent review. Hydrodynamics is an effective theory that describes 
the long-wavelength limit of the microscopic dynamics of a system. 
As a macroscopic theory which describes the space-time evolution of 
the energy-momentum tensor, it is much less involved than 
microscopic descriptions such as kinetic theory. In order to study 
the hydrodynamic evolution of a system, it is natural to first 
employ the equations of ideal hydrodynamics. However, ideal 
hydrodynamics is based on the unrealistic assumption of local 
thermodynamic equilibrium which results in isentropic evolution. 
Moreover, since the quantum mechanical uncertainty principle 
provides a lower bound on the shear viscosity to entropy density 
ratio \cite{Danielewicz:1984ww,Kovtun:2004de}, the dissipative 
effects can not be ignored.

Eckart \cite{Eckart:1940zz} and Landau and Lifshitz \cite{Landau} 
were the first to formulate a relativistic theory of dissipative 
hydrodynamics, each with a different choice for the definition of 
hydrodynamic four-velocity. These theories are based on the 
assumption that the entropy four-current is linear in dissipative 
quantities and hence they are also known as first-order theories of 
dissipative fluids. The resulting equations for the dissipative 
quantities are essentially the relativistic analogue of the 
Navier-Stokes equations. However, the resulting equations of motion 
lead to parabolic differential equations which suffer from 
acausality and numerical instability. In order to rectify the 
undesirable features of first-order theories, extended theories of 
dissipative fluids were introduced by Grad \cite{Grad}, M\"uller 
\cite{Muller:1967zza} and Israel and Stewart \cite{Israel:1979wp}. 
These theories are based on the assumption that the entropy 
four-current contains terms quadratic in the dissipative fluxes and 
therefore are also known as second-order theories. The resulting 
equations of motion are hyperbolic in nature which preserves 
causality \cite{Huovinen:2008te} but may not guarantee stability. 

Second-order Israel-Stewart (IS) hydrodynamics has been quite 
successful in explaining a wide range of collective phenomena 
observed in ultra-relativistic heavy-ion collisions \cite 
{Heinz:2013th}. Despite its successes, the formulation of IS theory 
is based on certain approximations and assumptions. For instance, 
the original IS theory employs an arbitrary choice of the second 
moment of the Boltzmann equation to obtain the equations of motion 
for the dissipative currents \cite{Israel:1979wp}. Another 
assumption inherent in IS theory is the use of Grad's 14-moment 
approximation for the non-equilibrium distribution function \cite 
{Grad,Israel:1979wp}. Moreover, the IS theory is a second-order 
theory which neglects contributions from higher-order terms in the 
entropy four-current. It is thus of interest to extend the 
second-order theory beyond its present scope and determine the 
associated transport coefficients for a hydrodynamic system.

In a non-equilibrium system, the presence of thermodynamic gradients 
results in thermodynamic forces which in turn gives rise to various 
transport phenomena. Therefore transport coefficients such as 
viscosity, diffusivity and conductivity, are important to 
characterize the dynamics of a system. Precise knowledge of these 
transport coefficients and associated length and time scales is 
necessary in comparing observables with theoretical predictions. In 
order to calculate these transport coefficients from the underlying 
kinetic theory, it is convenient to first determine the 
non-equilibrium single particle phase-space distribution function 
$f(x,p)$. When the system is close to local thermodynamic 
equilibrium, two most commonly used methods to determine the form of 
$f(x,p)$ are (1) Grad's 14-moment approximation \cite{Grad} and (2) 
the Chapman-Enskog expansion \cite{Chapman}. Although both methods 
involve expanding $f(x,p)$ around its equilibrium value $f_0(x,p)$, 
in Refs.~\cite{Jaiswal:2013npa,Jaiswal:2013vta,Jaiswal:2014isa} it 
was shown that the Chapman-Enskog method in the relaxation-time 
approximation (RTA) gives better agreement with microscopic 
Boltzmann simulations as well as exact solutions of the RTA 
Boltzmann equation. Consistent derivation of the form of the 
dissipative equations and accurate determination of the associated 
transport coefficients is still an active research area \cite 
{Jaiswal:2013npa,Jaiswal:2013vta,Jaiswal:2014isa,Muronga:2003ta, 
Denicol:2010xn,York:2008rr,Denicol:2012cn,Jaiswal:2012qm, 
Jaiswal:2013fc,Bhalerao:2013aha,Bhalerao:2013pza,Martinez:2010sc, 
Martinez:2012tu,Florkowski:2013lza,Bazow:2013ifa,El:2009vj, 
Muronga:2010zz,Muronga:2014yua,Denicol:2014mca,Florkowski:2010cf, 
Florkowski:2014sfa}.

In this paper, we derive a new expression for the entropy 
four-current up to third-order in dissipative fluxes by employing 
the Chapman-Enskog expansion for the non-equilibrium distribution 
function. Although third-order expressions for entropy four-current 
have been derived using Grad's 14-moment approximation \cite 
{El:2009vj,Muronga:2010zz,Muronga:2014yua}, we present here, for the 
first time, the derivation using the Chapman-Enskog method. We show 
that unlike second-order and third-order results from Grad's method, 
there is a non-vanishing entropy flux (projection of the entropy 
four-current orthogonal to the fluid four-velocity) in our 
expression for the entropy four-current. We demonstrate the 
significance of the present derivation in the special case of a 
system undergoing boost-invariant Bjorken expansion. We show that 
compared to the Grad's method, the Chapman-Enskog method is able to 
reproduce better the exact solution of Boltzmann equation \cite 
{Florkowski:2013lza,Baym:1984np} as well as the BAMPS results \cite 
{El:2009vj,Xu:2004mz}.

\section{Iterative solution of the Boltzmann equation}

Evolution of the single particle phase-space distribution function, 
$f(x,p)$, is governed by the Boltzmann equation. In the absence of 
collisions, the particles propagate along geodesics which implies 
that $f(x,p)$ does not change along geodesics. Therefore for a 
geodesic parametrized by an affine parameter $\Lambda$, we have 
$df/d\Lambda=0$. When collisions are present, particles will no 
longer move along geodesics leading to a non-vanishing $df/d\Lambda$. 
Hence, in general, one can write the Boltzmann equation as
\begin{equation}\label{GRBE}
p^\mu\partial_\mu f + F^\mu\partial_\mu^{(p)}f =  {\cal C}[f],
\end{equation}
where $p^\mu$ is the particle four-momentum, $F^\mu$ is the external 
force felt by the particles and ${\cal C}[f]$ is the collision 
functional. 

In absence of any external forces and using the relaxation-time 
approximation for the collision term, Eq.~(\ref{GRBE}) can be 
rewritten as \cite{Anderson_Witting}
\begin{equation}\label{RBERTA}
p^\mu\partial_\mu f  =  -(u\cdot p)\frac{\delta f}{\tau_R},
\end{equation}
where $u\cdot p \equiv u_\mu p^\mu$, $\tau_R$ is the relaxation time 
and $\delta f\equiv f-f_0$ is the non-equilibrium part of the 
distribution function, $f_0$ being the equilibrium distribution 
function. In the following, we consider only classical massless 
particles obeying the Boltzmann statistics at vanishing chemical 
potential, i.e., $f_0=\exp(-\beta\,u\cdot p)$, where $\beta\equiv1/T$
is the inverse temperature.

Equation (\ref{RBERTA}) can be solved iteratively to obtain a 
Chapman-Enskog like expansion for the non-equilibrium part of the 
distribution function in powers of space-time gradients \cite
{Chapman,Romatschke:2011qp} 
\begin{equation}\label{CEE} 
\delta f= \delta f^{(1)} + \delta f^{(2)} + \delta f^{(3)} + \cdots, 
\end{equation}
where $\delta f^{(1)}$ is first-order in derivatives, $\delta 
f^{(2)}$ is second-order and so on. To first- and second-order in 
derivatives, one obtains
\begin{align}
\delta f^{(1)} &= -\frac{\tau_R}{u\!\cdot\! p} \, p^\mu \partial_\mu f_0, \label{FOC} \\
\delta f^{(2)} &= \frac{\tau_R}{u\!\cdot\! p}p^\mu p^\nu\partial_\mu\Big(\frac{\tau_R}{u\!\cdot\! p} \partial_\nu f_0\Big). \label{SOC}
\end{align}
Derivation of hydrodynamic evolution equations for dissipative 
quantities within the framework of kinetic theory requires the form 
of $\delta f$ to be specified. In the following, we employ Eq.~(\ref
{CEE}) along with Eqs.~(\ref{FOC}) and (\ref{SOC}) to specify the 
non-equilibrium distribution function.

\section{Relativistic viscous hydrodynamics}

The hydrodynamic evolution of a relativistic system, in absence of 
any conserved charges, is governed by the conservation equation of 
the energy-momentum tensor. In terms of the phase-space distribution 
function and hydrodynamic variables, the conserved energy-momentum 
tensor can be expressed as \cite{deGroot}
\begin{align}\label{NTD}
T^{\mu\nu} &= \!\int\! dp \ p^\mu p^\nu\, f(x,p) = \epsilon u^\mu u^\nu 
- P\Delta ^{\mu \nu} + \pi^{\mu\nu},
\end{align}
where $dp\equiv g d{\bf p}/[(2 \pi)^3|\bf p|]$, $g$ being the 
degeneracy factor, $\epsilon$, $P$ and $\pi^{\mu\nu}$ are 
respectively energy density, pressure and the shear stress tensor. 
For a system of massless particles the bulk viscous pressure 
vanishes. The projection operator $\Delta^{\mu\nu}\equiv 
g^{\mu\nu}-u^\mu u^\nu$ is orthogonal to the hydrodynamic 
four-velocity $u^\mu$ defined in the Landau frame: $T^{\mu\nu} 
u_\nu=\epsilon u^\mu$. We consider the metric tensor to be 
Minkowskian, i.e., $g^{\mu\nu}\equiv\mathrm{diag}(+,-,-,-)$.

The evolution equations for $\epsilon$ and $u^\mu$ are obtained 
from the fundamental energy-momentum conservation, $\partial_\mu 
T^{\mu\nu} =0$, 
\begin{align}
\dot\epsilon + (\epsilon+P)\theta - \pi^{\mu\nu}\sigma_{\mu\nu} &= 0,  \label{evol01}\\
(\epsilon+P)\dot u^\alpha - \nabla^\alpha P + \Delta^\alpha_\nu \partial_\mu \pi^{\mu\nu}  &= 0.\label{evol02} 
\end{align}
We use the standard notation $\dot A\equiv u^\mu\partial_\mu A$ for 
co-moving derivative, $\theta\equiv \partial_\mu u^\mu$ for the 
expansion scalar, $\sigma_{\mu\nu}\equiv(\nabla_{\mu}u_{\nu} + 
\nabla_{\nu}u_{\mu})/2-(\theta/3)\Delta_{\mu\nu}$ for velocity 
stress tensor and $\nabla^\alpha\equiv\Delta^{\mu\alpha} 
\partial_\mu$ for space-like derivative. In the conformal limit, the 
energy density and pressure are related through $\epsilon=3P\propto 
\beta^{-4}$, where the inverse temperature $\beta\equiv1/T$ is 
defined using the equilibrium matching condition $\epsilon = 
\epsilon_0$. In this limit the derivatives of $\beta$ can be 
obtained using Eqs.~(\ref{evol01}) and (\ref{evol02})
\begin{align}
\dot\beta &= \frac{\beta}{3}\theta - \frac{\beta}{12P}\pi^{\rho\gamma}\sigma_{\rho\gamma}, \label{evol1}\\
\nabla^\alpha\beta &= \!-\beta\dot u^\alpha - \frac{\beta}{4P} \Delta^\alpha_\rho \partial_\gamma \pi^{\rho\gamma}. \label{evol2}
\end{align}
In the following, we will employ the above identities to derive the 
form of dissipative corrections to the distribution function as well 
as the evolution equation for shear stress tensor. 

In terms of $\delta f$, shear stress tensor ($\pi^{\mu\nu}$) can be 
expressed as
\begin{align}\label{FSE}
\pi^{\mu\nu} &= \Delta^{\mu\nu}_{\alpha\beta} \int dp \, p^\alpha p^\beta\, \delta f,
\end{align}
where $\Delta^{\mu\nu}_{\alpha\beta}\equiv 
\Delta^{\mu}_{(\alpha}\Delta^{\nu}_{\beta)} - 
(1/3)\Delta^{\mu\nu}\Delta_{\alpha\beta}$ is a traceless symmetric 
projection operator orthogonal to $u^\mu$. The first-order 
expression for shear stress tensor can be obtained from Eq.~(\ref
{FSE}) using $\delta f = \delta f^{(1)}$ from Eq.~(\ref {FOC}),
\begin{align}
\pi^{\mu\nu} &= \Delta^{\mu\nu}_{\alpha\beta}\int dp \ p^\alpha p^\beta 
\left(-\frac{\tau_R}{u\!\cdot\! p} \, p^\mu \partial_\mu\, f_0\right) . \label{FOSE}
\end{align}
Using Eqs.~(\ref{evol1}) and (\ref{evol2}), and retaining terms 
which are first-order in gradients, the integrals in the above 
equation reduce to
\begin{equation}\label{FOE}
\pi^{\mu\nu} = 2\tau_R\beta_\pi\sigma^{\mu\nu}, 
\end{equation}
where $\beta_\pi = 4P/5$.

In order to obtain higher-order evolution equations, we consider the 
co-moving derivative of Eq.~(\ref{FSE}),
\begin{equation}
\dot\pi^{\langle\mu\nu\rangle} = \Delta^{\mu\nu}_{\alpha\beta} \int dp\, p^\alpha p^\beta\, \delta\dot f, \label{SSE}
\end{equation}
where we have used the notation $A^{\langle\mu\nu\rangle}\equiv 
\Delta^{\mu\nu}_{\alpha\beta}A^{\alpha\beta}$ for traceless 
symmetric projection orthogonal to $u^{\mu}$. The co-moving 
derivative of the non-equilibrium part of the distribution function, 
$\delta\dot f$, can be obtained by rewriting Eq.~(\ref{RBERTA}) in 
the form \cite{Denicol:2010xn}
\begin{equation}\label{DFD}
\delta\dot f = -\dot f_0 - \frac{1}{u\!\cdot\! p}p^\gamma\nabla_\gamma f - \frac{\delta f}{\tau_R}.
\end{equation}
Using this expression for $\delta\dot f$ in Eq.~(\ref{SSE}), we obtain
\begin{equation}\label{SOSE}
\dot\pi^{\langle\mu\nu\rangle} + \frac{\pi^{\mu\nu}}{\tau_R} = 
- \Delta^{\mu\nu}_{\alpha\beta} \!\int\! \frac{dp}{u\!\cdot\! p} \, p^\alpha p^\beta p^\gamma\nabla_\gamma f. 
\end{equation}
From the above equation, we can conclude that the shear relaxation 
time $\tau_\pi$ is equal to the Boltzmann relaxation time $\tau_R$. 
A comparison of the first-order evolution equation, Eq.~(\ref 
{FOE}), with the relativistic Navier-Stokes equation, $\pi^{\mu\nu} =
2\eta\sigma^{\mu\nu}$, results in $\tau_\pi=\eta/\beta_\pi$ for the 
shear relaxation time.

To derive the second-order evolution equation for $\pi^{\mu\nu}$, we 
substitute $\delta f^{(1)}$ from Eq.~(\ref{FOC}) in Eq.~(\ref{SOSE}) 
and use Eqs.~(\ref{evol1}) and (\ref{evol2}) for derivatives of $\beta$.
We finally obtain \cite{Jaiswal:2013npa}
\begin{equation}\label{SOSHEAR}
\dot{\pi}^{\langle\mu\nu\rangle} \!+ \frac{\pi^{\mu\nu}}{\tau_\pi}\!= 
2\beta_{\pi}\sigma^{\mu\nu}
\!+2\pi_\gamma^{\langle\mu}\omega^{\nu\rangle\gamma}
\!-\frac{10}{7}\pi_\gamma^{\langle\mu}\sigma^{\nu\rangle\gamma} 
\!-\frac{4}{3}\pi^{\mu\nu}\theta,
\end{equation}
where $\omega^{\mu\nu}\equiv(\nabla^\mu u^\nu-\nabla^\nu u^\mu)/2$ 
is the vorticity tensor. We observe that by employing the above 
equation, $\delta f$ in Eqs.~(\ref{CEE})-(\ref{SOC}) can be 
expressed in terms of derivatives of hydrodynamic variables up to 
second order. To this end, we write
\begin{equation}\label{SOVC}
\delta f = f_0\phi = f_0\left(\phi_1 + \phi_2\right) + {\cal O}(\delta^3),
\end{equation}
where $\phi_1$ and $\phi_2$ are first- and second-order corrections, 
respectively, and are calculated to be
\begin{align}
\phi_1 =\, & \frac{\beta}{2\beta_\pi(u\!\cdot\!p)}\, p^\alpha p^\beta \pi_{\alpha\beta}, \label{phi1}\\
\phi_2 =\, & \frac{\beta}{\beta_\pi} \bigg[\frac{5}{14\beta_\pi (u\!\cdot\!p)}\, p^\alpha p^\beta \pi^\gamma_\alpha\, \pi_{\beta\gamma}
-\frac{\tau_\pi}{u\!\cdot\!p}\, p^\alpha p^\beta \pi^\gamma_\alpha\, \omega_{\beta\gamma} \nonumber\\
&-\frac{(u\!\cdot\!p)}{70\beta_\pi}\, \pi^{\alpha\beta}\pi_{\alpha\beta}
+\frac{6\tau_\pi}{5}\, p^\alpha\dot u^\beta\pi_{\alpha\beta}
-\frac{\tau_\pi}{5}\, p^\alpha\!\left(\nabla^\beta\pi_{\alpha\beta}\!\right) \nonumber\\
&-\frac{\tau_\pi}{2(u\!\cdot\!p)^2}\, p^\alpha p^\beta p^\gamma\!\left(\nabla_\gamma\pi_{\alpha\beta}\!\right)
+\frac{3\tau_\pi}{(u\!\cdot\!p)^2}\, p^\alpha p^\beta p^\gamma \pi_{\alpha\beta}\dot u_\gamma \nonumber\\
&-\frac{\tau_\pi}{3(u\!\cdot\!p)}\, p^\alpha p^\beta \pi_{\alpha\beta}\theta
+\frac{\beta+(u\!\cdot\!p)^{-1}}{4(u\!\cdot\!p)^2\beta_\pi}\left(p^\alpha p^\beta \pi_{\alpha\beta}\right)^2\bigg]. \label{phi2}
\end{align}
Here we have used Eqs.~(\ref{evol1}), (\ref{evol2}) and (\ref 
{SOSHEAR}) to substitute for the derivatives of $\beta$ and $u^\mu$. 
We observe that the form of $\phi_1$ and $\phi_2$ in Eqs.~(\ref 
{phi1})-(\ref{phi2}) satisfies the matching condition $\epsilon 
=\epsilon_0$ and the Landau frame definition $u_\nu T^{\mu \nu} = 
\epsilon u^\mu$ and is also consistent with Eq.~(\ref{FSE}) for the 
definition of the shear stress tensor \cite{Bhalerao:2013pza}. Note 
that the form of $\delta f$ obtained by Denicol {\it et. al.} \cite 
{Denicol:2012cn}, where they generalize the 14-moment method to 
include all terms in the moment expansion, also satisfies these 
conditions. However, unlike Eq.~(\ref{phi2}), the $\delta f$ 
obtained in Ref.~\cite{Denicol:2012cn} is linear in hydrodynamic 
gradients.
 
The third-order evolution equation for the shear stress tensor can 
also be derived by substituting $f=f_0(1+\phi_1+\phi_2)$ in Eq.~(\ref
{SOSE}). After straightforward but tedious algebra, we obtain 
\cite {Jaiswal:2013vta}
\begin{align}\label{TOSHEAR}
\dot{\pi}^{\langle\mu\nu\rangle} =& -\frac{\pi^{\mu\nu}}{\tau_\pi}
+2\beta_\pi\sigma^{\mu\nu}
+2\pi_{\gamma}^{\langle\mu}\omega^{\nu\rangle\gamma}
-\frac{10}{7}\pi_\gamma^{\langle\mu}\sigma^{\nu\rangle\gamma}  \nonumber\\
&-\frac{4}{3}\pi^{\mu\nu}\theta
+\frac{25}{7\beta_\pi}\pi^{\rho\langle\mu}\omega^{\nu\rangle\gamma}\pi_{\rho\gamma}
-\frac{1}{3\beta_\pi}\pi_\gamma^{\langle\mu}\pi^{\nu\rangle\gamma}\theta \nonumber\\
&-\frac{38}{245\beta_\pi}\pi^{\mu\nu}\pi^{\rho\gamma}\sigma_{\rho\gamma}
-\frac{22}{49\beta_\pi}\pi^{\rho\langle\mu}\pi^{\nu\rangle\gamma}\sigma_{\rho\gamma} \nonumber\\
&-\frac{24}{35}\nabla^{\langle\mu}\left(\pi^{\nu\rangle\gamma}\dot u_\gamma\tau_\pi\right)
+\frac{4}{35}\nabla^{\langle\mu}\left(\tau_\pi\nabla_\gamma\pi^{\nu\rangle\gamma}\right) \nonumber\\
&-\frac{2}{7}\nabla_{\gamma}\left(\tau_\pi\nabla^{\langle\mu}\pi^{\nu\rangle\gamma}\right)
+\frac{12}{7}\nabla_{\gamma}\left(\tau_\pi\dot u^{\langle\mu}\pi^{\nu\rangle\gamma}\right) \nonumber\\
&-\frac{1}{7}\nabla_{\gamma}\left(\tau_\pi\nabla^{\gamma}\pi^{\langle\mu\nu\rangle}\right)
+\frac{6}{7}\nabla_{\gamma}\left(\tau_\pi\dot u^{\gamma}\pi^{\langle\mu\nu\rangle}\right) \nonumber\\
&-\frac{2}{7}\tau_\pi\omega^{\rho\langle\mu}\omega^{\nu\rangle\gamma}\pi_{\rho\gamma}
-\frac{2}{7}\tau_\pi\pi^{\rho\langle\mu}\omega^{\nu\rangle\gamma}\omega_{\rho\gamma} \nonumber\\
&-\frac{10}{63}\tau_\pi\pi^{\mu\nu}\theta^2
+\frac{26}{21}\tau_\pi\pi_\gamma^{\langle\mu}\omega^{\nu\rangle\gamma}\theta.
\end{align}
We compare the above equation with that obtained in Ref. \cite 
{El:2009vj} using Grad's 14-moment approximation,
\begin{align}\label{TOEF}
\dot\pi^{\langle\mu\nu\rangle} =& -\frac{\pi^{\mu\nu}}{\tau_\pi'} + 2\beta_\pi'\sigma^{\mu\nu}
-\frac{4}{3}\pi^{\mu\nu}\theta 
+ \frac{5}{36\beta_\pi'}\pi^{\mu\nu}\pi^{\rho\gamma}\sigma_{\rho\gamma} \nonumber \\
&-\frac{16}{9\beta_\pi'}\pi^{\langle\mu}_{\gamma}\pi^{\nu\rangle\gamma}\theta,
\end{align}
where $\beta_\pi'=2P/3$ and $\tau_\pi'=\eta/\beta_\pi'$. Note that 
the right-hand side of the above equation contains one second-order 
and two third-order terms compared to three second-order and 
fourteen third-order terms obtained in Eq.~(\ref{TOSHEAR}).

\section{Entropy four-current}

A well established framework for the study of thermalization 
processes in a system begins from the observation that thermal 
equilibrium corresponds to the state of maximum entropy. The 
interpretation of how entropy is generated in any process depends on 
the theoretical and conceptual framework in which the processes that 
lead to thermalization are described. For instance, in a 
relativistic system, local entropy generation is given by the 
divergence of the entropy four-current. For kinetic theory, the 
expression for entropy four-current generalized from the Boltzmann's 
H-function is given by
\begin{equation}\label{EFC}
S^\mu = -\int dp ~p^\mu f \left(\ln f - 1\right).
\end{equation}
For a system which is close to local thermodynamic equilibrium, 
$f=f_0(1+\phi)$, where $\phi\ll 1$, we obtain an expression for the 
non-equilibrium entropy four-current up to third-order in $\phi$ as
\begin{equation}\label{TEFC}
S^\mu = s_0 u^\mu - \int dp ~p^\mu f_0 \left(\frac{\phi^2}{2}-\frac{\phi^3}{6}\right),
\end{equation}
where $s_0=\beta(\epsilon+P)$ is the equilibrium entropy density. 
For $\phi=\phi_1+\phi_2$, we have
\begin{equation}\label{TOEFC}
S^\mu = s_0 u^\mu - \int dp ~p^\mu f_0 \left(\frac{\phi_1^2}{2}+\phi_1\phi_2-\frac{\phi_1^3}{6}\right),
\end{equation}
where we have ignored terms which are higher than third-order in 
derivative expansion.

Substituting $\phi_1$ and $\phi_2$ from Eqs.~(\ref{phi1}) and (\ref
{phi2}) and performing the integrations, we get
\begin{align}\label{TOEFCF}
S^\mu =&~ s_0 u^\mu - \frac{\beta}{4\beta_\pi}\pi^{\alpha\beta}\pi_{\alpha\beta}u^\mu
-\frac{5\beta}{42\beta_\pi^2}\pi_{\alpha\gamma}\pi^\gamma_\beta\pi^{\alpha\beta}u^\mu \nonumber\\
&~+\!\frac{\beta\tau_\pi}{7\beta_\pi}\bigg[\frac{18}{5}\dot u^\rho \pi_{\rho\gamma}\pi^{\mu\gamma} 
\!+\frac{2}{5}\pi^{\mu\gamma}\nabla^\rho\pi_{\rho\gamma}
\!-\frac{1}{2}\pi^{\alpha\beta}\nabla^\mu\pi_{\alpha\beta}\nonumber\\
&~+3\dot u^\mu\pi_{\alpha\beta}\pi^{\alpha\beta}
-\pi^{\alpha\gamma}\Delta^{\mu\rho}\nabla_\alpha\pi_{\rho\gamma}\bigg],
\end{align}
where we recall that $\beta_\pi=4P/5$. We compare our above result 
with that obtained using Grad's 14-moment approximation \cite
{El:2009vj},
\begin{equation}\label{TOEFCG}
S'^\mu = s_0 u^\mu - \frac{\beta}{4\beta_\pi'}\pi^{\alpha\beta}\pi_{\alpha\beta}u^\mu
-\frac{2\beta}{9\beta_\pi'^{2}}\pi_{\alpha\gamma}\pi^\gamma_\beta\pi^{\alpha\beta}u^\mu,
\end{equation}
where $\beta_\pi'=2P/3$.\footnote{We note that the factor $2/9$ in 
Eq.~(\ref{TOEFCG}) is four times larger than that obtained in Ref. 
\cite{Muronga:2014yua}, despite the fact that both methods employ 
Grad's 14-moment approximation.} The entropy density, $s\equiv u_\mu 
S^\mu$, for the two cases is given by
\begin{align}
s =&~ s_0 - \frac{\beta}{4\beta_\pi}\pi^{\alpha\beta}\pi_{\alpha\beta}
-\frac{5\beta}{42\beta_\pi^2}\pi_{\alpha\gamma}\pi^\gamma_\beta\pi^{\alpha\beta}, \label{TOEDCE}\\
s' =&~ s_0 - \frac{\beta}{4\beta_\pi'}\pi^{\alpha\beta}\pi_{\alpha\beta}
-\frac{2\beta}{9\beta_\pi'^{2}}\pi_{\alpha\gamma}\pi^\gamma_\beta\pi^{\alpha\beta}, \label{TOEDG}
\end{align}
whereas the entropy flux, $S^{\langle\mu\rangle}\equiv\Delta^\mu_\nu 
S^\nu$, in the two cases reduce to
\begin{align}
S^{\langle\mu\rangle} =&~ \frac{\beta\tau_\pi}{7\beta_\pi}\bigg[\frac{18}{5}\dot u^\rho \pi_{\rho\gamma}\pi^{\mu\gamma} 
\!+\frac{2}{5}\pi^{\mu\gamma}\nabla^\rho\pi_{\rho\gamma}
\!-\frac{1}{2}\pi^{\alpha\beta}\nabla^\mu\pi_{\alpha\beta}\nonumber\\
&+3\dot u^\mu\pi_{\alpha\beta}\pi^{\alpha\beta}
-\pi^{\alpha\gamma}\Delta^{\mu\rho}\nabla_\alpha\pi_{\rho\gamma}\bigg], \label{TOEFLCE}\\
S'^{\langle\mu\rangle} =&~ 0. \label{TOEFLG}
\end{align}
We observe that even for vanishing bulk viscosity and dissipative 
charge current, Chapman-Enskog method leads to non-vanishing entropy 
flux as opposed to the method based on Grad's 14-moment 
approximation. This may be attributed to the fact that the entropy 
four-current obtained in the Chapman-Enskog method contains terms 
proportional to acceleration $\dot u^\mu$ and gradient of shear 
stress tensor $\nabla^\mu\pi_{\alpha\beta}$. Both these quantities, 
as well as $\pi_{\alpha\beta}$, are orthogonal to the fluid 
four-velocity $u^\mu$ and therefore their combination results in 
non-vanishing entropy flux. Note that for a system with vanishing 
entropy flux, the entropy four-flow is entirely due to the flow of 
entropy density. In the case of Chapman-Enskog method, the 
non-vanishing entropy flux implies that the entropy density of the 
system should be lower than in the case of vanishing entropy flux 
(Grad's method).

\section{Numerical results and discussion}

\begin{figure}[t]
 \begin{center}
  \scalebox{.248}{\includegraphics{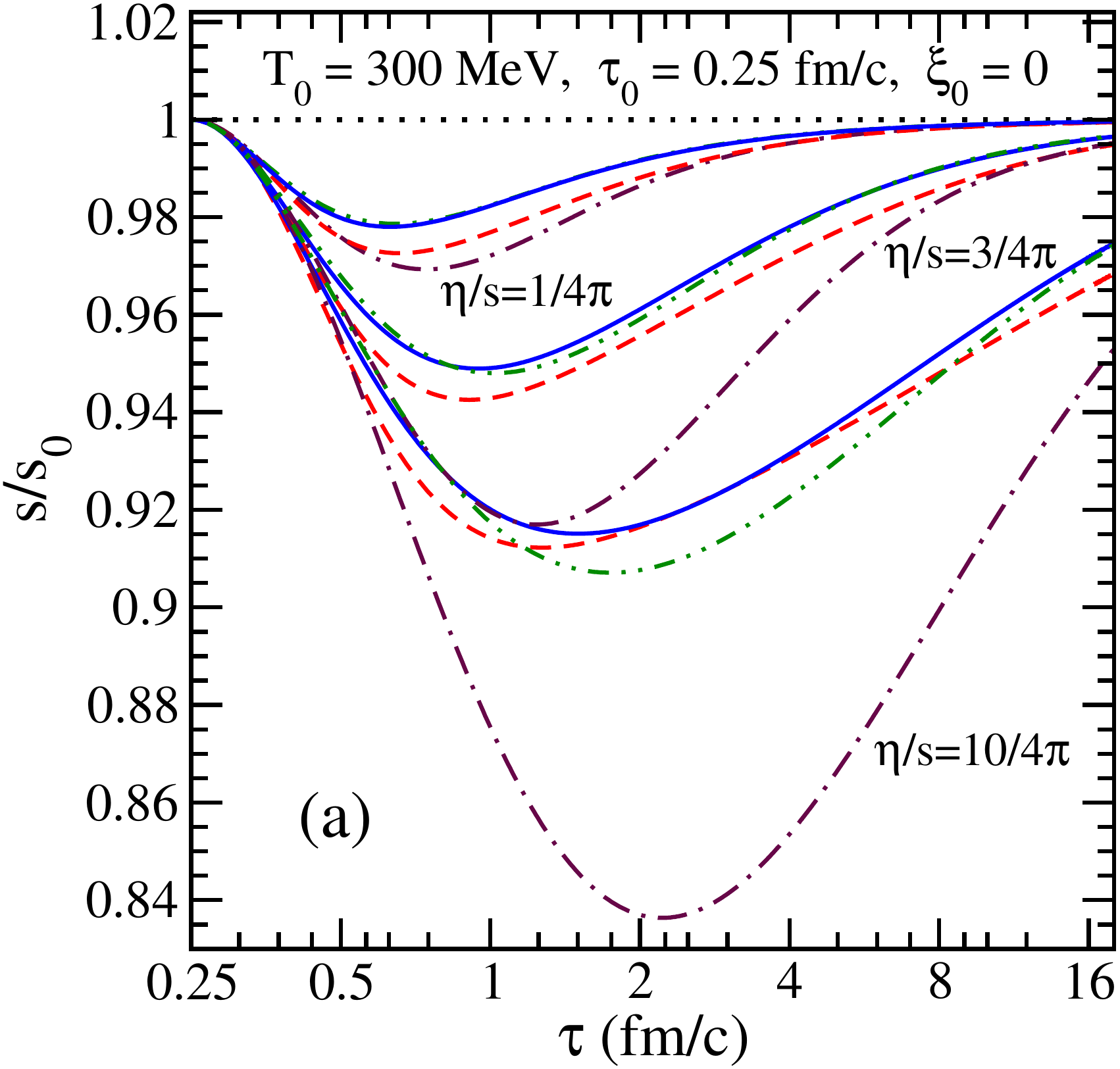}}\hfil
  \scalebox{.248}{\includegraphics{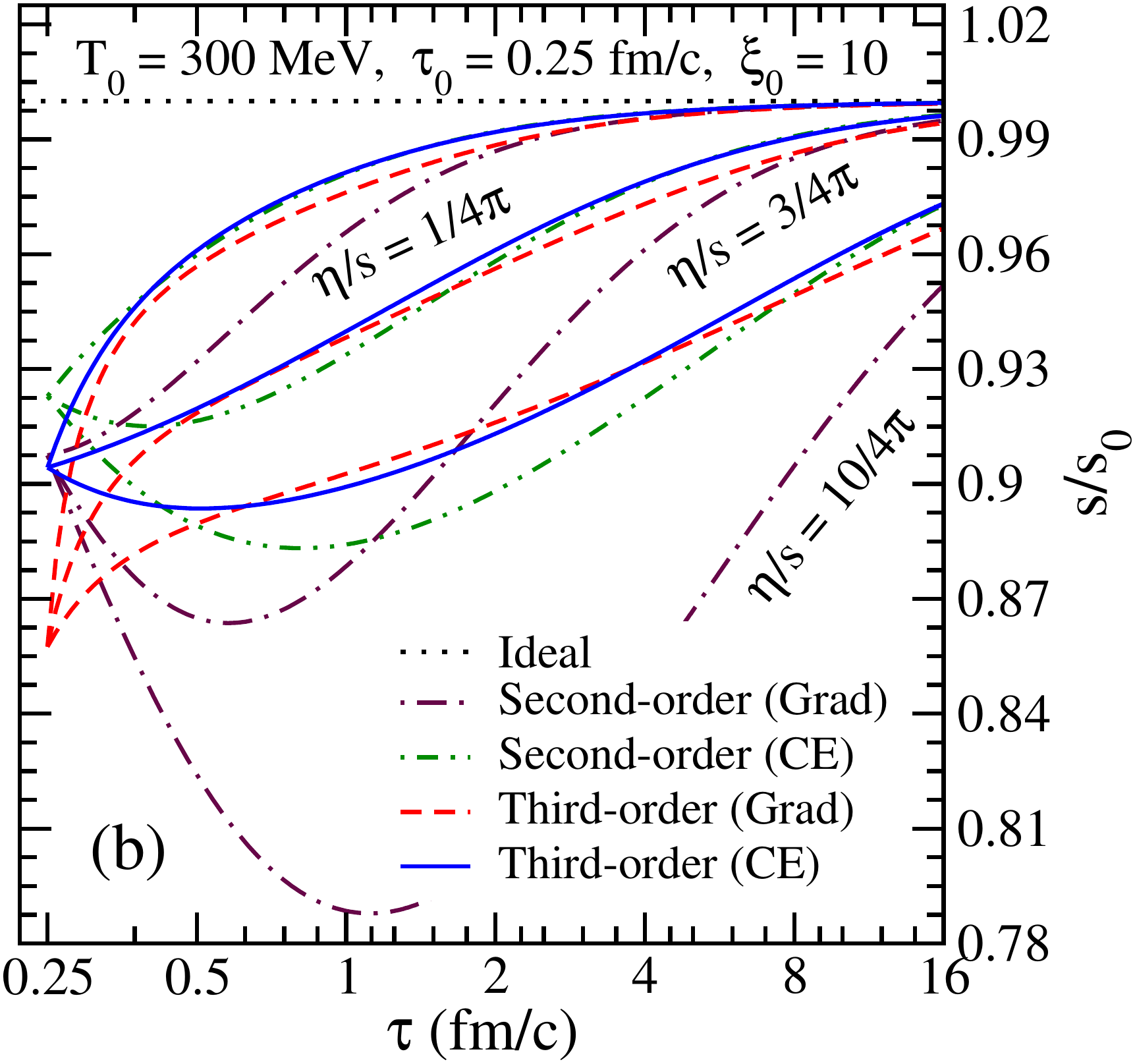}}
 \end{center}
 \vspace{-0.6cm}
 \caption{(Color online) Time evolution of $s/s_0$ obtained 
  using ideal hydrodynamics (black dotted lines), second-order 
  Grad's approximation (maroon dashed-dotted lines), second-order 
  Chapman-Enskog method (green dashed-dotted-dotted lines), 
  third-order Grad's approximation (red dashed lines), and 
  third-order Chapman-Enskog method (blue solid lines) for initial 
  temperature $T_0=300$ MeV at initial time $\tau_0=0.25$ fm/c, 
  various $\eta/s$ and for (a): isotropic initial pressure 
  configuration $\xi_0=0$ and (b): anisotropic pressure 
  configuration $\xi_0=10$.}
 \label{Entropy_300}
\end{figure}

For a transversely homogeneous and purely-longitudinal 
boost-invariant Bjorken expansion of a system \cite{Bjorken:1982qr}, 
all scalar functions of space and time depend only on the proper 
time $\tau=\sqrt{t^2-z^2}$. In the Milne coordinate system 
$(\tau,x,y,\eta_s)$, where $\eta_s=\tanh^{-1}(z/t)$, the 
hydrodynamic four-velocity becomes $u^\mu=(1,0,0,0)$. In this 
scenario, $\omega^{\mu\nu}=\dot u^\mu=\nabla^\mu\tau_\pi=0$, $\theta 
= 1/\tau$, $\sigma^{\eta_s\eta_s} = -2/(3\tau^3)$ and only the 
$\eta_s\eta_s$ component of Eq.~(\ref{TOSHEAR}) survives. Defining 
$\pi\equiv-\tau^2\pi^{\eta_s\eta_s}$, we obtain
\begin{align}
\dot\pi^{\langle\eta_s\eta_s\rangle} &= -\frac{1}{\tau^2}\frac{d\pi}{d\tau},  &
\pi^{\langle\eta_s}_{\gamma}\sigma^{\eta_s\rangle\gamma} &= -\frac{\pi}{3\tau^3}, \nonumber\\
\pi^{\langle\eta_s}_{\gamma}\pi^{\eta_s\rangle\gamma} &= -\frac{\pi^2}{2\tau^2},  & 
\pi^{\rho\gamma}\sigma_{\rho\gamma} &= \pi/\tau, \nonumber\\ 
\pi^{\rho\langle\eta_s}\pi^{\eta_s\rangle\gamma}\sigma_{\rho\gamma} &= -\frac{\pi^2}{2\tau^3},  &
\nabla^{\langle\eta_s}\nabla_{\gamma}\pi^{\eta_s\rangle\gamma} &= \frac{2\pi}{3\tau^4}, \nonumber\\ 
\nabla_{\gamma}\nabla^{\langle\eta_s}\pi^{\eta_s\rangle\gamma} &= \frac{4\pi}{3\tau^4},  &
\nabla^2\pi^{\langle\eta_s\eta_s\rangle} &= \frac{4\pi}{3\tau^4}.
\label{identity}
\end{align} 
Using the above results, evolution of $\epsilon$ and $\pi$ from 
Eqs.~(\ref{evol01}) and (\ref{TOSHEAR}) becomes
\begin{align}
\frac{d\epsilon}{d\tau} &= -\frac{1}{\tau}\left(\epsilon + P  -\pi\right), \label{BED} \\
\frac{d\pi}{d\tau} &= - \frac{\pi}{\tau_\pi} + \beta_\pi\frac{4}{3\tau} - \lambda\frac{\pi}{\tau} 
- \chi\frac{\pi^2}{\beta_\pi\tau}. \label{Bshear}
\end{align}
The terms with coefficient $\lambda$ and $\chi$ in the above equation 
contains corrections due to second-order and third-order terms, 
respectively. In order to rewrite some of the third-order 
contributions in the form $\pi^2/(\beta_\pi\tau)$, the first-order 
expression for shear pressure, $\pi=4\beta_\pi\tau_\pi/3\tau$, has 
been used. The transport coefficients in Eq.~(\ref{Bshear}) 
simplify to
\begin{equation}\label{BTC}
\tau_\pi = \frac{\eta}{\beta_\pi}, \quad \beta_\pi = \frac{4P}{5}, \quad \lambda = \frac{38}{21}, \quad \chi = \frac{72}{245}.
\end{equation}
While the form of Eq.~(\ref{TOEF}), obtained using Grad's 14-moment 
approximation, is identical to Eq.~(\ref {Bshear}) in the Bjorken 
case, the transport coefficients reduce to
\begin{equation}\label{BTCE}
\tau_\pi' = \frac{\eta}{\beta_\pi'}, \quad \beta_\pi' = \frac{2P}{3}, \quad \lambda' = \frac{4}{3}, \quad \chi' = \frac{3}{4}.
\end{equation}

\begin{figure}[t]
 \begin{center}
  \scalebox{.248}{\includegraphics{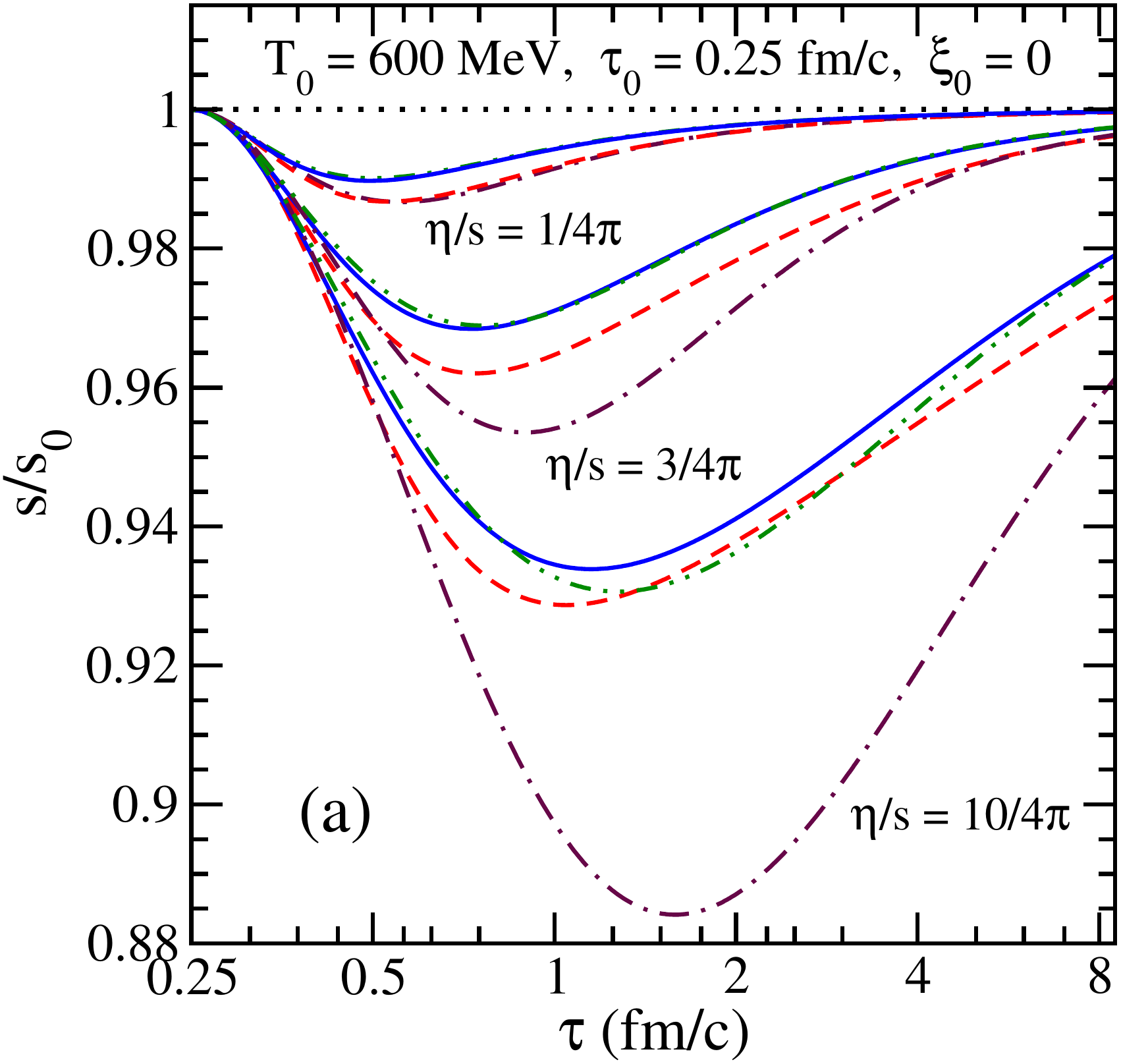}}\hfil
  \scalebox{.248}{\includegraphics{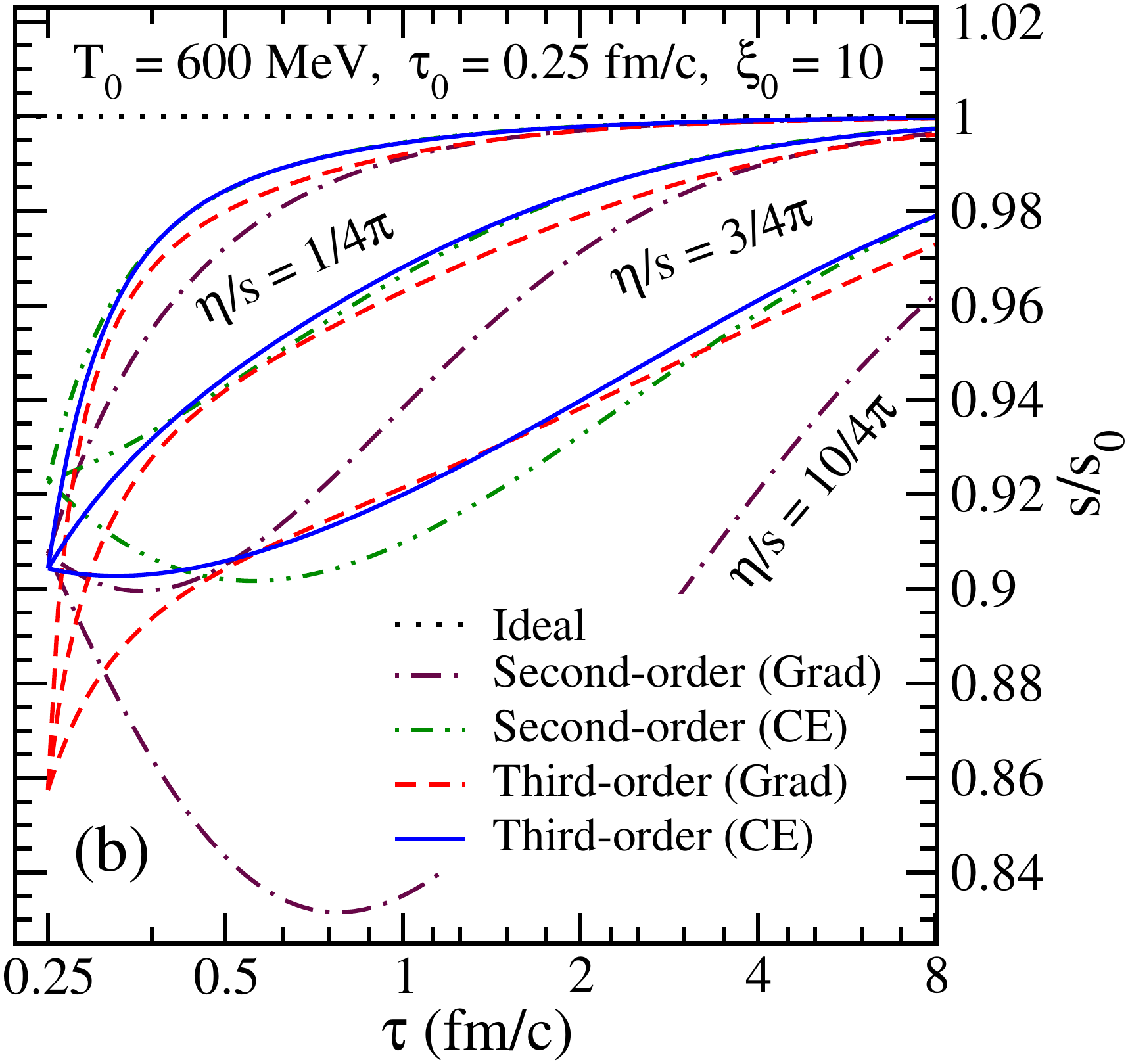}}
 \end{center}
 \vspace{-0.6cm}
 \caption{(Color online) Same as Fig.~\ref{Entropy_300} except 
 here we take $T_0=600$ MeV.}
 \label{Entropy_600}
\end{figure}

\begin{figure}[t]
 \begin{center}
  \scalebox{.249}{\includegraphics{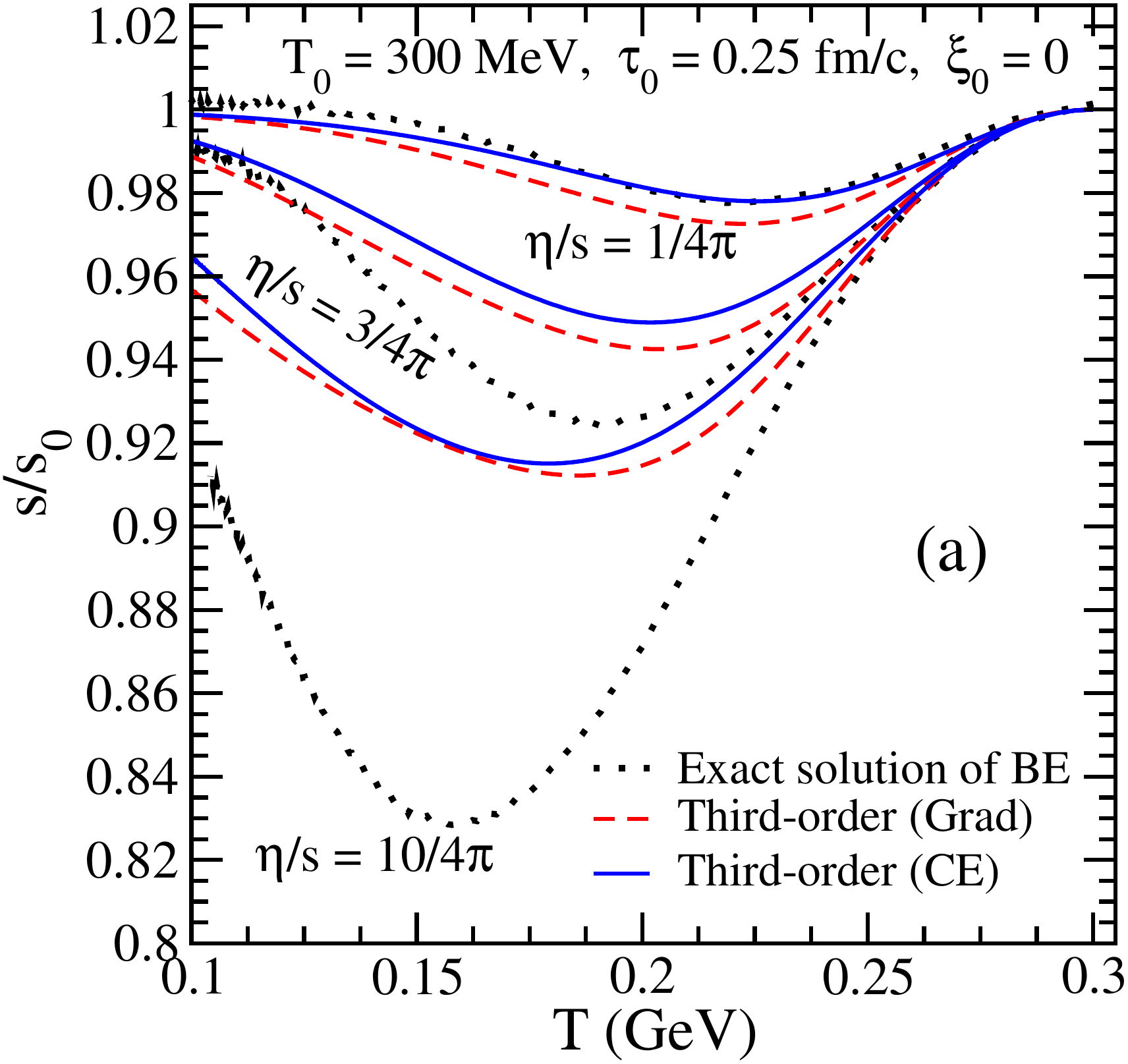}}\hfil
  \scalebox{.249}{\includegraphics{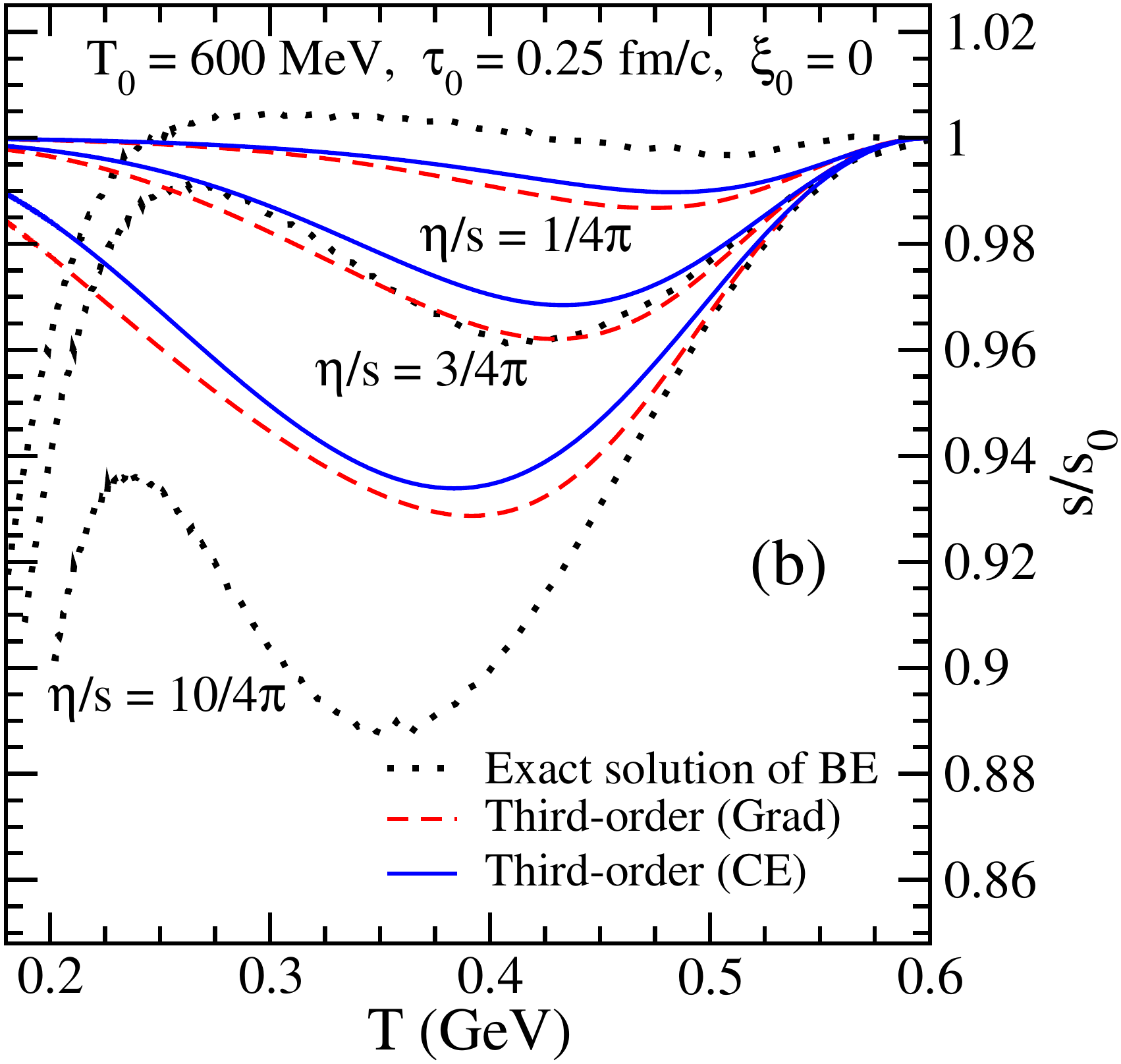}}
 \end{center}
 \vspace{-0.6cm}
 \caption{(Color online) Temperature dependence of $s/s_0$ obtained 
  using exact solution of the Boltzmann equation (black dotted 
  lines), third-order evolution equations from Grad's approximation 
  (red dashed lines), and third-order equations from Chapman-Enskog 
  method (blue solid lines) for various $\eta/s$ and for (a): 
  initial temperature $T_0=300$ MeV and (b): $T_0=600$ MeV, at 
  initial time $\tau_0=0.25$ fm/c and isotropic initial pressure 
  configuration $\xi_0=0$.}
 \label{Entropy_T}
\end{figure}

We solve Eqs.~(\ref{BED}) and (\ref{Bshear}) simultaneously assuming 
two different initial temperatures, $T_0=300$ MeV and $T_0=600$ MeV, 
at the initial proper time $\tau_0=0.25$ fm/c. The initial pressure 
configurations are determined by the anisotropy parameter $\xi$ 
which is related to the average transverse and longitudinal momentum 
in the local rest frame via the relation $\xi=\frac{1}{2} \langle 
p_T^2\rangle/\langle p_L^2\rangle-1$ \cite{Martinez:2010sc}. We 
solve for two different initial pressure configurations: $\xi_0=0$ 
corresponding to an isotropic pressure configuration $\pi_0=0$, and 
$\xi_0=10$ corresponding to $\pi_0= 87$ MeV/fm$^3$ for $T_0=300$ MeV 
and $\pi_0=1386$ MeV/fm$^3$ for $T_0=600$ MeV. For comparison, we 
also solve Eqs.~(\ref {BED}) and (\ref{Bshear}) with transport 
coefficients obtained using the Grad's 14-moment method \cite 
{El:2009vj}.

\begin{figure}[t]
 \begin{center}
  \scalebox{.26}{\includegraphics{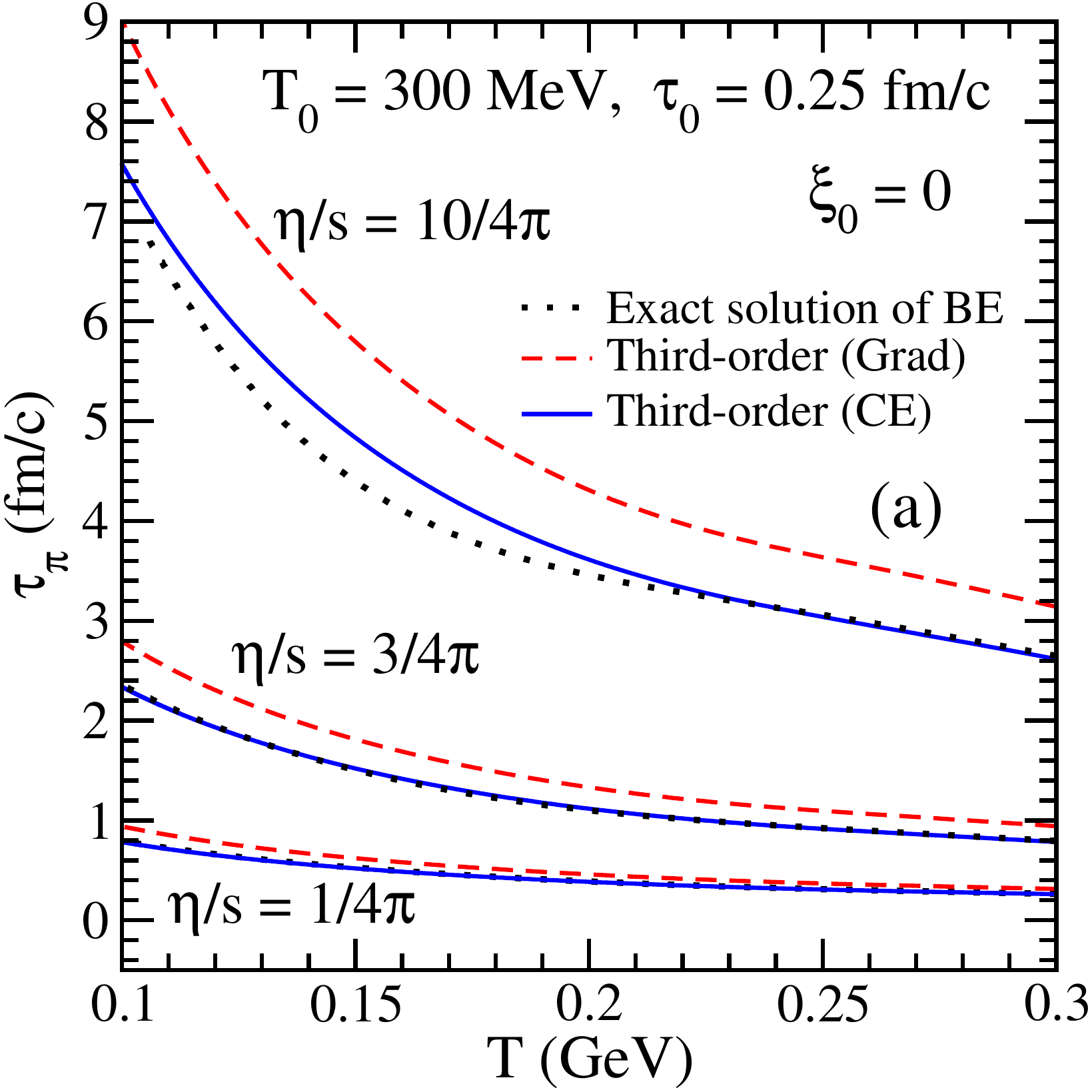}}\hfil
  \scalebox{.26}{\includegraphics{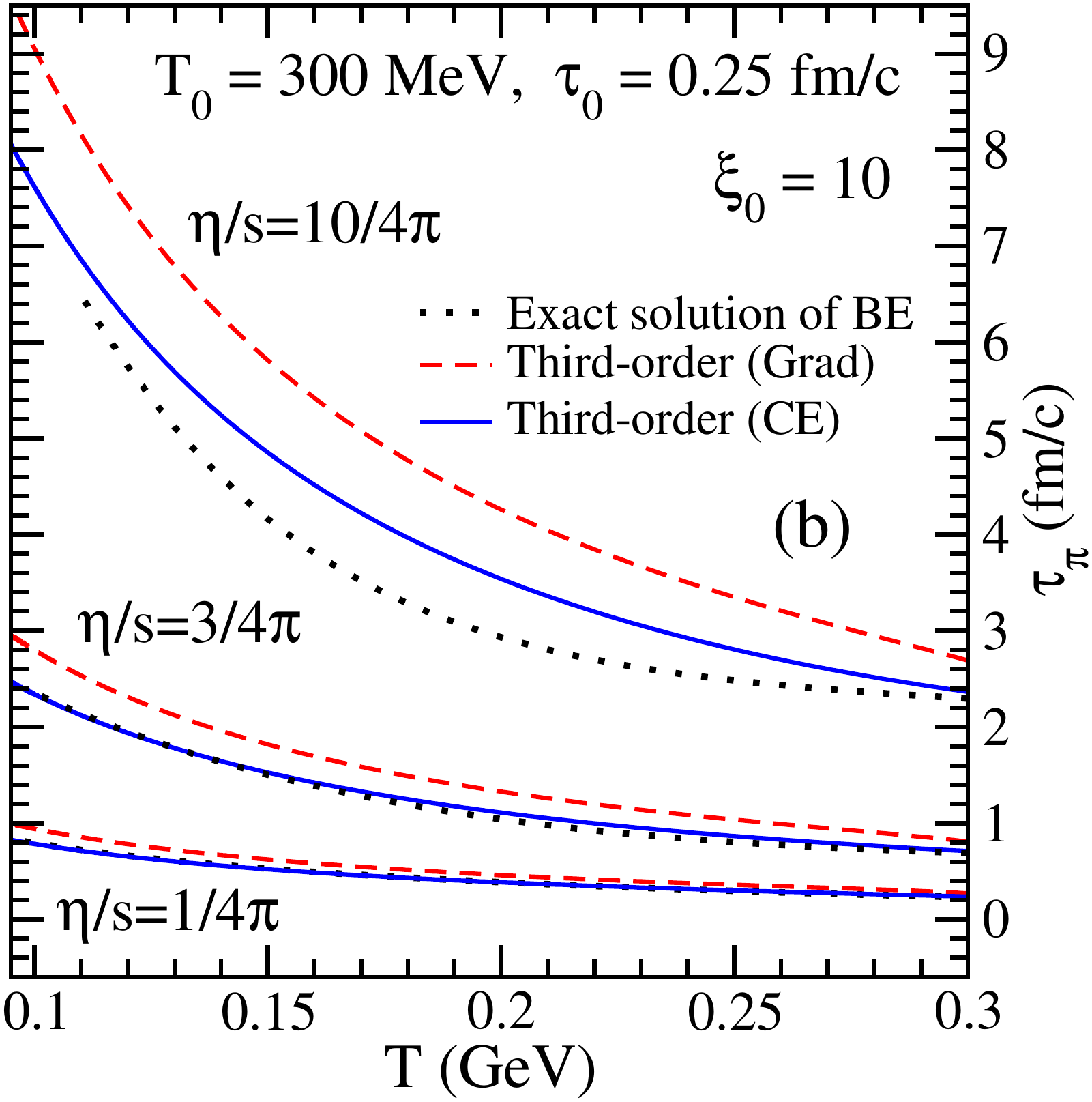}}
 \end{center}
 \vspace{-0.6cm}
 \caption{(Color online) Temperature dependence of the shear 
  relaxation time, $\tau_\pi$, obtained using exact solution of the 
  Boltzmann equation (black dotted lines), third-order evolution 
  equations from Grad's approximation (red dashed lines), and 
  third-order equations from Chapman-Enskog method (blue solid 
  lines) for initial temperature $T_0=300$ MeV at initial time 
  $\tau_0=0.25$ fm/c, various $\eta/s$ and for (a): isotropic 
  initial pressure configuration $\xi_0=0$ and (b): anisotropic 
  pressure configuration, $\xi_0=10$.}
 \label{Tau_300}
\end{figure}

\begin{figure}[t]
 \begin{center}
  \scalebox{.26}{\includegraphics{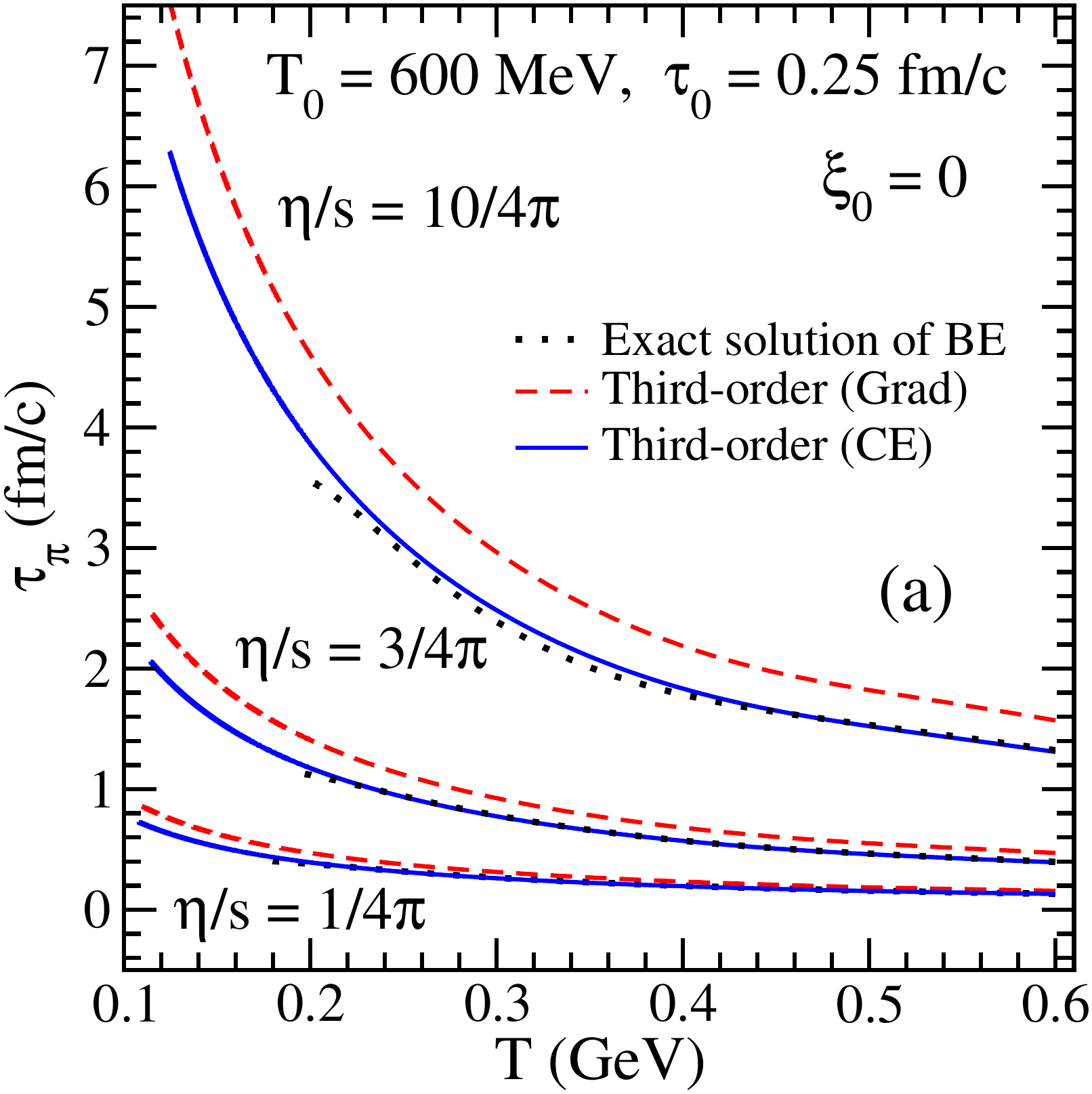}}\hfil
  \scalebox{.26}{\includegraphics{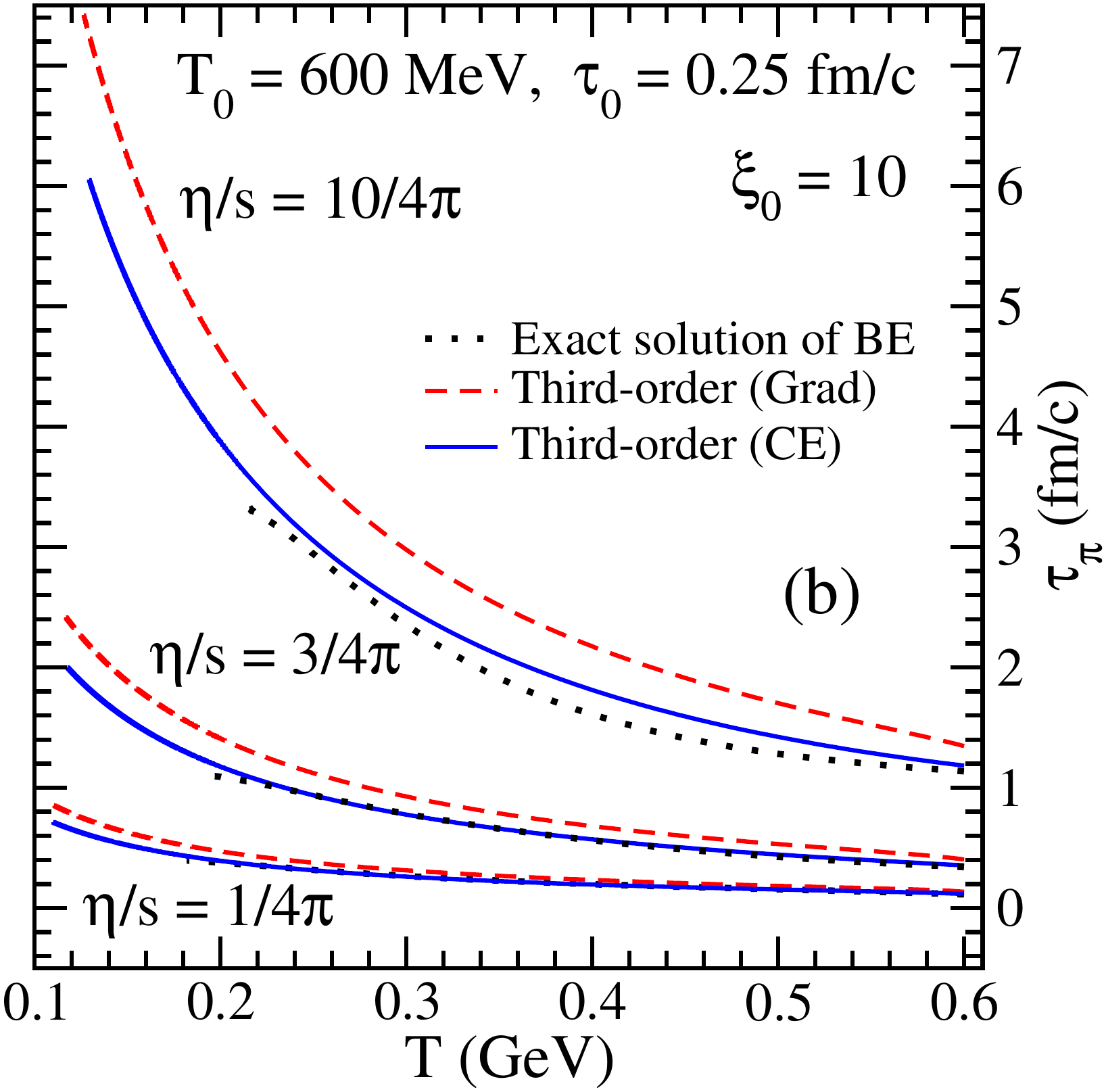}}
 \end{center}
 \vspace{-0.6cm}
 \caption{(Color online) Same as Fig.~\ref{Tau_300} except 
 here we take $T_0=600$ MeV.}
 \label{Tau_600}
\end{figure}

\begin{figure}[t]
 \begin{center}
  \scalebox{.255}{\includegraphics{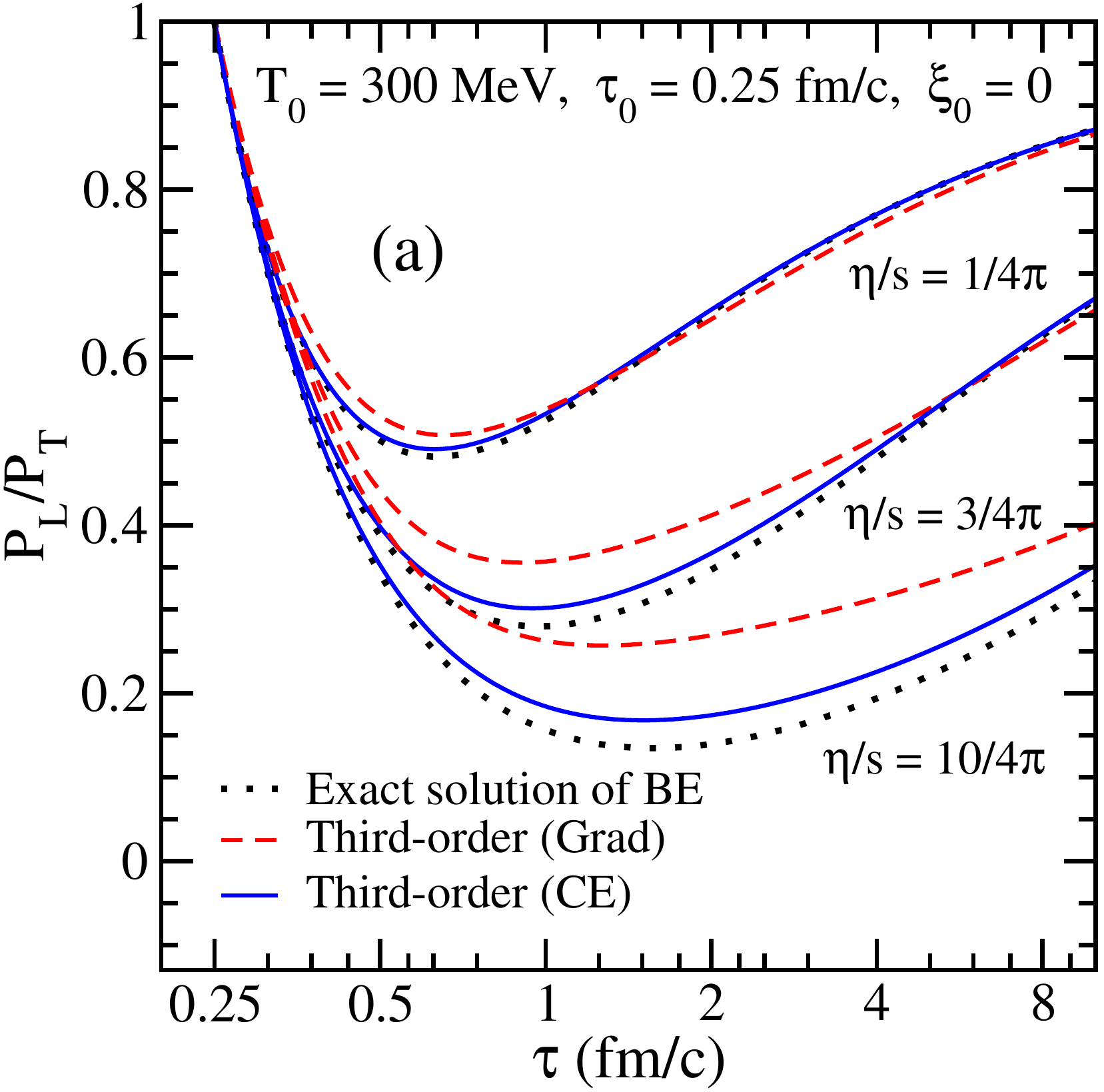}}\hfil
  \scalebox{.255}{\includegraphics{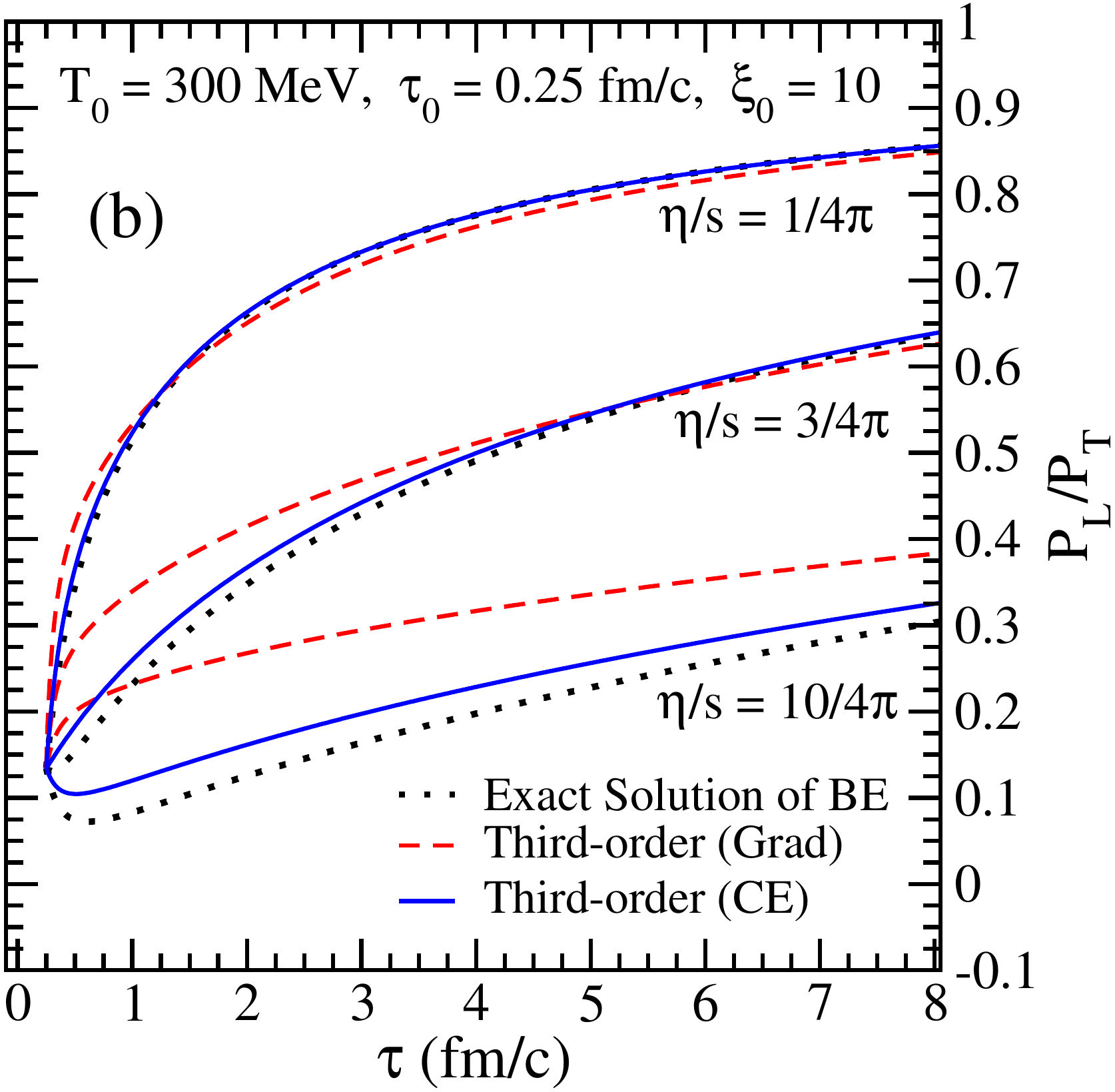}}
 \end{center}
 \vspace{-0.6cm}
 \caption{(Color online) Time evolution of $P_L/P_T$ obtained 
  using exact solution of the Boltzmann equation (black dotted 
  lines), third-order evolution equations from Grad's approximation 
  (red dashed lines), and third-order equations from Chapman-Enskog 
  method (blue solid lines) for initial temperature $T_0=300$ MeV at 
  initial time $\tau_0=0.25$ fm/c, various $\eta/s$ and for (a): 
  isotropic initial pressure configuration $\xi_0=0$ and (b): 
  anisotropic pressure configuration, $\xi_0=10$.}
 \label{PLPT_300}
\end{figure}

\begin{figure}[t]
 \begin{center}
  \scalebox{.257}{\includegraphics{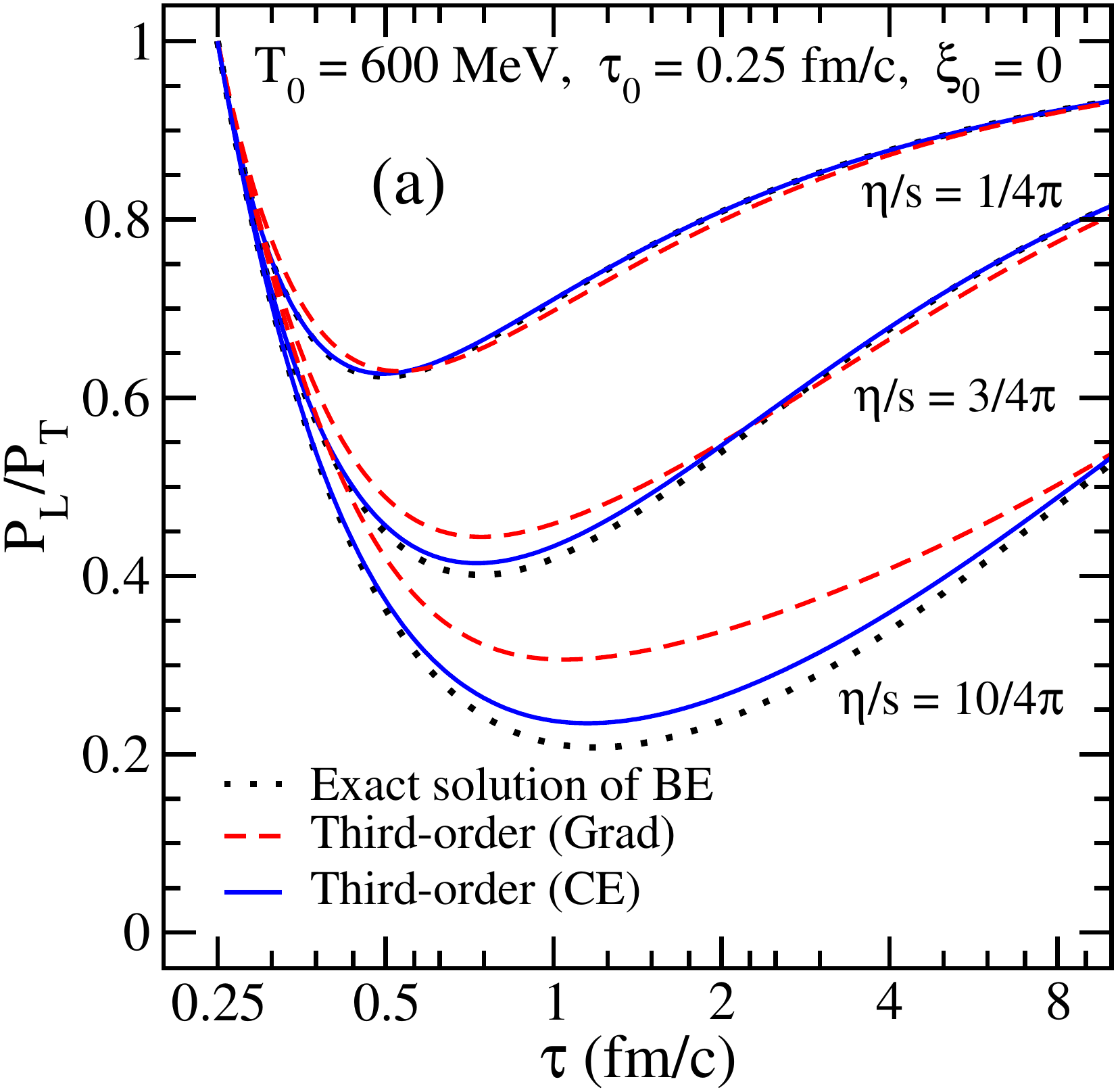}}\hfil
  \scalebox{.257}{\includegraphics{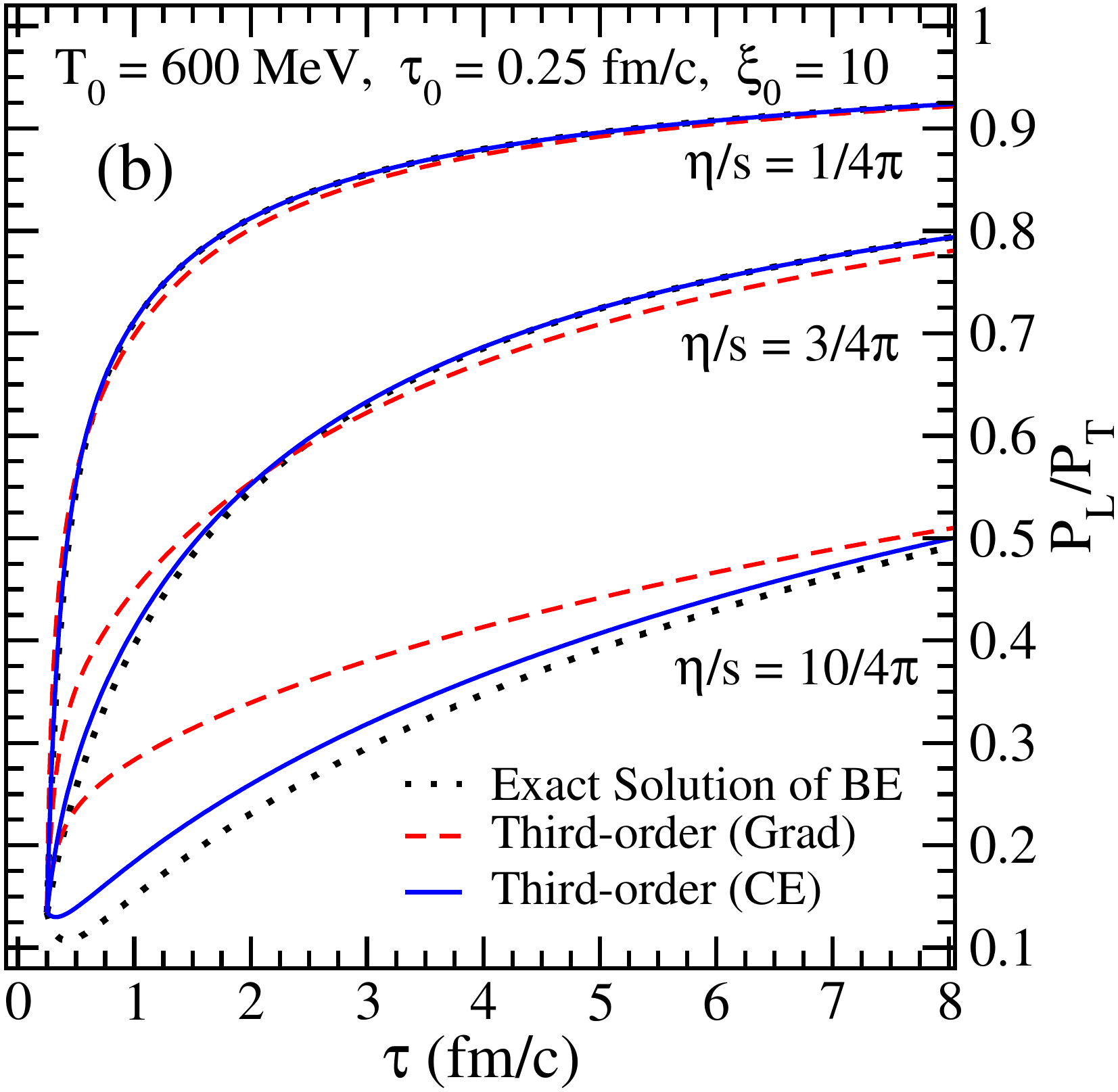}}
 \end{center}
 \vspace{-0.6cm}
 \caption{(Color online) Same as Fig.~\ref{PLPT_300} except 
 here we take $T_0=600$ MeV.}
 \label{PLPT_600}
\end{figure}

In Figs.~\ref{Entropy_300} and \ref{Entropy_600} we show the 
proper-time evolution of the entropy density scaled by its 
equilibrium value, $s/s_0$, obtained using ideal hydrodynamics 
(black dotted lines), second-order Grad's approximation (maroon 
dashed-dotted lines), second-order Chapman-Enskog (green 
dashed-dotted-dotted lines), third-order Grad's approximation (red 
dashed lines), and third-order Chapman-Enskog method (blue solid 
lines). Figure~\ref{Entropy_300} shows the case when initial 
temperature $T_0=300$ MeV, while Fig.~\ref{Entropy_600} shows the 
case that $T_0=600$ MeV. In both figures, panels (a) and (b) 
correspond to isotropic initial pressure configuration $\xi_0=0$ and 
anisotropic pressure configuration $\xi_0=10$, respectively, and the 
initial time $\tau_0=0.25$ fm/c.

In the left panels of Figs.~\ref{Entropy_300} and \ref{Entropy_600}, 
we see that $s/s_0$ shows a minimum indicating that the initial and 
final states of the system are close to equilibrium. In the 
intermediate stage, viscous evolution leads to significant deviation 
of the entropy density from its equilibrium value. Moreover, we also 
observe that for Grad's method, second-order results are highly 
sensitive to $\eta/s$ and third-order contribution is very large, 
especially for large $\eta/s$. On the other hand, Chapman-Enskog 
method shows less sensitivity to $\eta/s$ and has small third-order 
contribution indicating faster convergence compared to the Grad's 
method. From Figs.~\ref{Entropy_300} and \ref{Entropy_600}, we also 
observe that the entropy density attains its equilibrium value more 
rapidly for higher $T_0$ indicating that the system equilibrates 
faster for larger initial temperature. On the other hand, in 
Fig.~\ref{Entropy_T}, we see that both Grad's method (red dashed lines) 
and Chapman-Enskog method (blue solid lines) are unable to reproduce 
the temperature dependence of $s/s_0$ obtained using the exact 
solution of the Boltzmann equation \cite {Florkowski:2013lza, 
Baym:1984np} (black dotted lines) for $T_0=300$ MeV and $T_0=600$ MeV.

In Figs.~\ref{Tau_300} and \ref{Tau_600} we show the temperature 
dependence of the shear relaxation time, $\tau_\pi$, and in 
Figs.~\ref{PLPT_300} and \ref{PLPT_600} we show the proper time 
evolution of the pressure anisotropy, $P_L/P_T\equiv(P-\pi)/(P+\pi/2)$. 
The presented results correspond to exact solution of the Boltzmann 
equation (black dotted lines), third-order evolution equations from 
Grad's approximation (red dashed lines), and third-order equations 
from Chapman-Enskog method (blue solid lines). Figures~\ref{Tau_300} 
and \ref{PLPT_300} show the case when initial temperature $T_0=300$ 
MeV, while Figs.~\ref{Tau_600} and \ref{PLPT_600} show the case that 
$T_0=600$ MeV. In Figs.~\ref{Tau_300} -- \ref{PLPT_600}, panels (a) 
and (b) correspond to isotropic initial pressure configuration 
$\xi_0=0$ and anisotropic pressure configuration $\xi_0=10$, 
respectively, and the initial time $\tau_0=0.25$ fm/c. 

From Figs.~\ref {Tau_300} -- \ref{PLPT_600}, we see that results 
obtained using the Grad's method always overestimate the shear 
relaxation time and fails to reproduce the pressure anisotropy 
obtained by the exact solution of the Boltzmann equation \cite 
{Florkowski:2013lza,Baym:1984np}. On the other hand, the 
Chapman-Enskog method clearly shows a better agreement with the 
exact solution of the Boltzmann equation and appreciable differences 
are observed only for the case of $\eta/s=10/4\pi$. We note that 
although both methods fail to reproduce the temperature dependence 
of $s/s_0$ obtained using the exact solution of the Boltzmann 
equation (see Fig.~\ref {Entropy_T}), the evolution of pressure 
anisotropy and shear relaxation time are found to be similar. 
Therefore we may conclude that the evolution of hydrodynamic 
quantities are insensitive to small variations in $s/s_0$. These 
results also indicate that the Chapman-Enskog method is better 
suited than the Grad's method to capture the microscopic dynamics 
contained in the Boltzmann equation.

\begin{figure}[t]
\begin{center}
\includegraphics[scale=0.4]{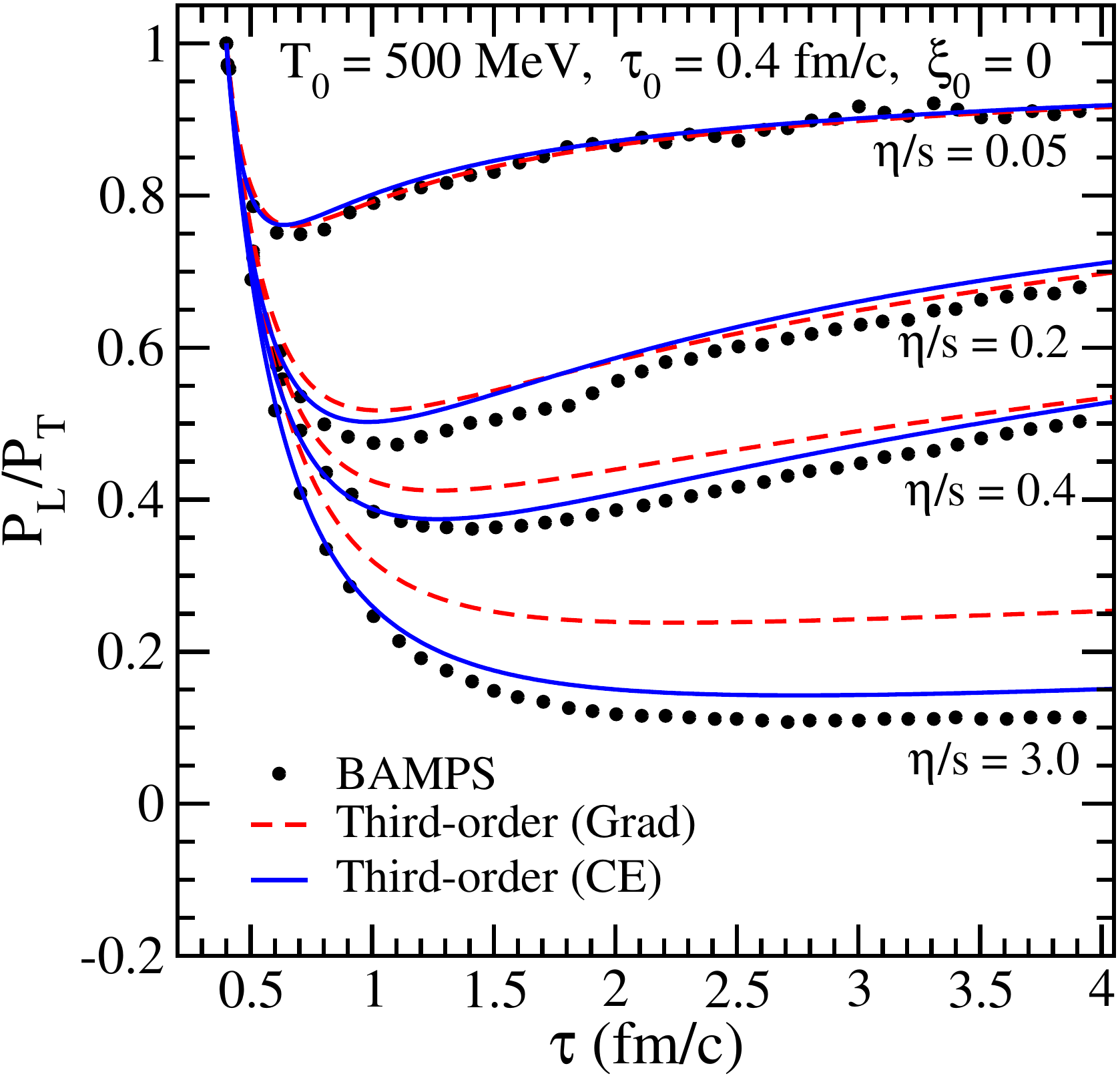}
\end{center}
\vspace{-0.4cm}
\caption{(Color online) Time evolution of $P_L/P_T$ obtained using 
  BAMPS (black dots), third-order evolution equations from Grad's 
  approximation (red dashed lines), and third-order equations from 
  Chapman-Enskog method (blue solid lines) for initial temperature 
  $T_0=500$ MeV at initial time $\tau_0=0.4$ fm/c, isotropic initial 
  pressure configuration $\xi_0=0$ and various $\eta/s$.} 
\label{PLPT}
\end{figure}

In order to compare our results with a transport model, the parton 
cascade BAMPS \cite{El:2009vj,Xu:2004mz}, we also solve Eqs.~(\ref 
{BED}) and (\ref{Bshear}) for $T_0=500$ MeV at $\tau_0=0.4$ fm/c and 
$\xi_0=0$. In Fig.~\ref{PLPT}, we show the proper time evolution of 
$P_L/P_T$ obtained using BAMPS (black dots), third-order evolution 
equations from Grad's approximation (red dashed lines), and 
third-order equations from Chapman-Enskog method (blue solid lines). 
We see that also in this case the $P_L/P_T$ obtained using the 
Chapman-Enskog method show better agreement with BAMPS results 
compared to Grad's method. This result confirms our previous 
observation that Chapman-Enskog method is better adapted than the 
Grad's method to capture the microphysics contained in the Boltzmann 
equation.

\section{Conclusions and outlook}

In this paper, we have employed the iterative solution of the 
Boltzmann equation in relaxation-time approximation to derive a new 
expression for the entropy four-current up to third order in 
gradient expansion. We found that unlike second-order and 
third-order entropy four-current obtained using Grad's method, there 
is a non-vanishing entropy flux in our expression even in the 
absence of bulk viscosity and dissipative charge current. Having 
obtained the full set of third-order evolution equations necessary 
to evolve the shear tensor, we then considered the special case of a 
transversally homogeneous and longitudinally boost-invariant system. 
In this particular case the Boltzmann equation in the 
relaxation-time approximation can be solved exactly \cite 
{Florkowski:2013lza,Baym:1984np}. Using this solution as a 
benchmark, we computed the entropy density, the shear relaxation 
time and pressure anisotropy using both the Chapman-Enskog method 
presented herein and the Grad's 14-moment method used in Ref.~\cite 
{El:2009vj}. We also compared the pressure anisotropy obtained using 
both the Chapman-Enskog method presented herein and the Grad's 
method with the results of the parton cascade BAMPS \cite 
{El:2009vj,Xu:2004mz}. We demonstrated that the Chapman-Enskog 
method is able to reproduce the exact solution of Boltzmann equation 
as well as the BAMPS results better than the Grad's method.

As a final remark, we note that the relaxation-time approximation 
for the collision term in the Boltzmann equation is based on the 
assumption that the collisions tend to restore the distribution 
function to its local equilibrium value exponentially. While it is 
true that the microscopic interactions of the constituent particles 
are not captured here, it is a reasonably good approximation to 
describe a system which is close to local thermodynamic equilibrium. 
Indeed, it was shown that for a purely gluonic system at weak 
coupling and hadron gas with large momenta, the Boltzmann equation 
in the relaxation-time approximation is a fairly accurate 
description \cite{Dusling:2009df}. Moreover, the experimentally 
observed and ideal hydrodynamic prediction of $1/\sqrt{m_T}$ scaling 
of the femtoscopic radii was found to be violated by including 
viscous corrections to the distribution function using Grad's method 
\cite{Teaney:2003kp}. It was shown later that this scaling can be 
restored by using the form of the non-equilibrium distribution 
function obtained using the Chapman-Enskog method \cite 
{Bhalerao:2013pza}. It has also been demonstrated recently that in 
contrast to the Grad's approximation, the renormalization-group 
method leads to similar expressions for the transport coefficients 
as given by the Chapman-Enskog method \cite{Tsumura:2013uma}. Hence 
we can conclude that the Boltzmann equation in the relaxation-time 
approximation can be applied quite successfully toward understanding 
the hydrodynamic behaviour of the strongly interacting matter formed 
in relativistic heavy-ion collisions.

At this juncture, we would like to clarify that we are using the 
exact solution of the Boltzmann equation in relaxation-time 
approximation \cite{Florkowski:2013lza,Baym:1984np} as a benchmark 
to compare different hydrodynamic formulations. We demand that the 
minimal requirement for a viable conformal hydrodynamic theory is 
that it should be able to describe the evolution of a viscous medium 
in this simple case. Moreover, we have demonstrated that the 
Chapman-Enskog method leads to a fairly good agreement with the 
BAMPS results \cite{El:2009vj,Xu:2004mz} which employs a more 
realistic collision kernel. Looking forward, it would be interesting 
to see if the third-order results derived herein could be extended 
to a system having bulk viscosity and dissipative charge current. 
Furthermore, since a large bulk viscosity might lead to an early 
onset of cavitation, it would therefore be instructive to see how 
the third-order transport coefficients could influence cavitation 
\cite {Jaiswal:2013fc}. In addition, from a phenomenological 
perspective, it would be interesting to determine the impact of the 
third-order evolution equations in higher-dimensional simulations. 
We leave these questions for future work.

\begin{acknowledgments}

A.J. was supported by the Frankfurt Institute for Advanced Studies 
(FIAS). R.R. was supported by Polish National Science Center Grant 
No. DEC-2012/07/D/ST2/02125.

\end{acknowledgments}


\begin{thebibliography}{99}

\bibitem{Heinz:2013th} 
  U.~Heinz and R.~Snellings,
  Ann.\ Rev.\ Nucl.\ Part.\ Sci.\  {\bf 63}, 123 (2013).

\bibitem{Danielewicz:1984ww} 
  P.~Danielewicz and M.~Gyulassy,
  Phys.\ Rev.\ D {\bf 31}, 53 (1985).
  
\bibitem{Kovtun:2004de} 
  P.~Kovtun, D.~T.~Son and A.~O.~Starinets,
  Phys.\ Rev.\ Lett.\  {\bf 94}, 111601 (2005).

\bibitem{Eckart:1940zz} 
  C.~Eckart,
  Phys.\ Rev.\  {\bf 58}, 267 (1940).

\bibitem{Landau} L.D. Landau and E.M. Lifshitz, {\it Fluid Mechanics}
  (Butterworth-Heinemann, Oxford, 1987).

  \bibitem{Grad}
H. Grad, Comm. Pure Appl. Math. {\bf 2}, 331 (1949).

\bibitem{Muller:1967zza} 
  I.~Muller,
  Z.\ Phys.\  {\bf 198}, 329 (1967).

\bibitem{Israel:1979wp} 
  W.~Israel and J.~M.~Stewart,
  Ann. Phys.\  {\bf 118}, 341 (1979).

\bibitem{Huovinen:2008te} 
  P.~Huovinen and D.~Molnar,
  Phys.\ Rev.\ C {\bf 79}, 014906 (2009).

\bibitem{Chapman}
S.~Chapman and T.~G.~Cowling, {\it The Mathematical Theory of Non-Uniform Gases}, 
(Cambridge University Press, Cambridge, 1970), 3rd ed. 

\bibitem{Jaiswal:2013npa} 
  A.~Jaiswal,
  Phys.\ Rev.\ C {\bf 87}, 051901 (2013); 
  arXiv:1408.0867 [nucl-th].
  
\bibitem{Jaiswal:2013vta} 
  A.~Jaiswal,
  Phys.\ Rev.\ C {\bf 88}, 021903 (2013); 
  Nucl.\ Phys.\ A {\bf 931}, 1205 (2014).
  
\bibitem{Jaiswal:2014isa} 
  A.~Jaiswal, R.~Ryblewski and M.~Strickland,
  Phys.\ Rev.\ C {\bf 90}, 044908 (2014).
  
\bibitem{Muronga:2003ta} 
  A.~Muronga,
  Phys.\ Rev.\ C {\bf 69}, 034903 (2004)

\bibitem{Denicol:2010xn} 
  G.~S.~Denicol, T.~Koide and D.~H.~Rischke,
  Phys.\ Rev.\ Lett.\  {\bf 105}, 162501 (2010).
  
\bibitem{York:2008rr} 
  M.~A.~York and G.~D.~Moore,
  Phys.\ Rev.\ D {\bf 79}, 054011 (2009).

\bibitem{Denicol:2012cn} 
  G.~S.~Denicol, H.~Niemi, E.~Molnar and D.~H.~Rischke,
  Phys.\ Rev.\ D {\bf 85}, 114047 (2012).

\bibitem{Jaiswal:2012qm} 
  A.~Jaiswal, R.~S.~Bhalerao and S.~Pal,
  Phys.\ Lett.\ B {\bf 720}, 347 (2013);
  J.\ Phys.\ Conf.\ Ser.\  {\bf 422}, 012003 (2013);
  arXiv:1303.1892 [nucl-th].

\bibitem{Jaiswal:2013fc} 
  A.~Jaiswal, R.~S.~Bhalerao and S.~Pal,
  Phys.\  Rev.\ C {\bf 87}, 021901(R) (2013).

\bibitem{Bhalerao:2013aha} 
  R.~S.~Bhalerao, A.~Jaiswal, S.~Pal and V.~Sreekanth,
  Phys.\ Rev.\ C {\bf 88}, 044911 (2013).

\bibitem{Bhalerao:2013pza} 
  R.~S.~Bhalerao, A.~Jaiswal, S.~Pal and V.~Sreekanth,
  Phys.\ Rev.\ C {\bf 89}, 054903 (2014).
  
\bibitem{Martinez:2010sc} 
  M.~Martinez and M.~Strickland,
  Nucl.\ Phys.\ A {\bf 848}, 183 (2010).
    
\bibitem{Florkowski:2010cf} 
  W.~Florkowski and R.~Ryblewski,
  Phys.\ Rev.\ C {\bf 83}, 034907 (2011).

\bibitem{Martinez:2012tu} 
  M.~Martinez, R.~Ryblewski and M.~Strickland,
  Phys.\ Rev.\ C {\bf 85}, 064913 (2012).

\bibitem{Florkowski:2013lza} 
  W.~Florkowski, R.~Ryblewski and M.~Strickland,
  Nucl.\ Phys.\ A {\bf 916}, (2013) 249;
  Phys.\ Rev.\ C {\bf 88}, (2013) 024903.

\bibitem{Bazow:2013ifa} 
  D.~Bazow, U.~W.~Heinz and M.~Strickland,
  Phys.\ Rev.\ C {\bf 90}, 054910 (2014).

\bibitem{El:2009vj} 
  A.~El, Z.~Xu and C.~Greiner,
  Phys.\ Rev.\ C {\bf 81}, 041901 (2010).
  
\bibitem{Muronga:2010zz} 
  A.~Muronga,
  J.\ Phys.\ G {\bf 37}, 094008 (2010).

\bibitem{Muronga:2014yua} 
  A.~Muronga,
  Acta Phys.\ Polon.\ Supp.\  {\bf 7}, 197 (2014).
  
\bibitem{Denicol:2014mca} 
  G.~S.~Denicol, W.~Florkowski, R.~Ryblewski and M.~Strickland,
  Phys.\ Rev.\ C {\bf 90}, 044905 (2014).
  
\bibitem{Florkowski:2014sfa} 
  W.~Florkowski, E.~Maksymiuk, R.~Ryblewski and M.~Strickland,
  Phys.\ Rev.\ C {\bf 89}, 054908 (2014).
  
\bibitem{Baym:1984np} 
  G.~Baym,
  Phys.\ Lett.\ B {\bf 138}, 18 (1984).
  
\bibitem{Xu:2004mz} 
  Z.~Xu and C.~Greiner,
  Phys.\ Rev.\ C {\bf 71}, 064901 (2005); {\bf 76}, 024911 (2007).
  
\bibitem{Anderson_Witting}
J.~L.~Anderson and H.~R.~Witting
Physica \textbf{74}, 466 (1974).

\bibitem{Romatschke:2011qp} 
  P.~Romatschke,
  Phys.\ Rev.\ D {\bf 85}, 065012 (2012).
  
\bibitem{deGroot}
S.R. de Groot, W.A. van Leeuwen, and Ch.G. van Weert, 
{\it Relativistic Kinetic Theory --- Principles and Applications} (North-Holland, Amsterdam, 1980).
  
\bibitem{Bjorken:1982qr} 
  J.~D.~Bjorken,
  Phys.\ Rev.\ D {\bf 27}, 140 (1983).
  
\bibitem{Dusling:2009df} 
  K.~Dusling, G.~D.~Moore and D.~Teaney,
  Phys.\ Rev.\ C {\bf 81}, 034907 (2010)
  
\bibitem{Teaney:2003kp} 
  D.~Teaney,
  Phys.\ Rev.\ C {\bf 68}, 034913 (2003)
  
\bibitem{Tsumura:2013uma} 
  K.~Tsumura and T.~Kunihiro,
  arXiv:1311.7059 [physics.flu-dyn].

\end{thebibliography}
\end{document}